\documentclass[sigconf, nonacm]{acmart}



\usepackage[linesnumbered,ruled,vlined]{algorithm2e}
\usepackage{subcaption}
\usepackage{tikz}
\usepackage{amsmath}
\usepackage{amssymb}
\usepackage{enumitem}
\usepackage{float}

\newcommand{\stitle}[1]{\vspace*{0.4em}\noindent{\bf #1.\/}}
\newcommand{\sstitle}[1]{\vspace*{0.4em}\noindent{\bf #1:\/}}
\newcommand{\trim}{\vspace{-2mm}}

\newcommand{\squishlist}{
 \begin{list}{$\bullet$}
 { \setlength{\itemsep}{0pt}
   \setlength{\parsep}{0pt}
   \setlength{\topsep}{0pt}
   \setlength{\partopsep}{0pt}
   \setlength{\leftmargin}{1.0em}
   \setlength{\labelwidth}{1em}
   \setlength{\labelsep}{0.6em}
 }
}
\newcommand{\squishend}{
 \end{list}
}

\newcommand\sysname{\textsf{Yi}}

\newcommand\tasklets{tasklets}

\newcommand\tasklet{tasklet}
\newcommand\Tasklet{Tasklet}
\newcommand*\circled[1]{\tikz[baseline=(char.base)]{\node[shape=circle,draw,inner sep=0pt, minimum size=8pt, line width=0.3pt] (char) {\fontsize{7}{0}\selectfont{#1}};}}

\newcommand*\circledcolor[3]{%
  \tikz[baseline=(char.base)]{%
    \node[
      shape=circle,
      draw=#1,           
      text=#2,           
      inner sep=0pt,
      minimum size=8pt,
      line width=0.3pt
    ] (char) {\fontsize{7}{0}\selectfont{#3}};%
  }%
}

\newcommand\vldbdoi{XX.XX/XXX.XX}
\newcommand\vldbpages{XXX-XXX}
\newcommand\vldbvolume{14}
\newcommand\vldbissue{1}
\newcommand\vldbyear{2020}
\newcommand\vldbauthors{\authors}
\newcommand\vldbtitle{\shorttitle} 
\newcommand\vldbavailabilityurl{https://anonymous.4open.science/r/Yi-VLDB-618}
\newcommand\vldbpagestyle{plain} 

\begin{document}

\title{Efficient and Effective In-place Graph-based Vector Index Updates}

\author{Haotian Liu$^*$, Yujun He$^*$, Bo Tang}
\affiliation{
  \institution{Southern University of Science and Technology}
}
\thanks{$^*$Equal contribution.}

\begin{abstract}
In the era of Large Language Models (LLMs), efficient vector updates are critical for capturing real-time information from rapidly evolving data.
However, it is not trivial to process frequent vector insert and delete updates and maintain a high recall of the search results simultaneously.
Specifically, the cluster-based vector indexing methods have high update throughput but low search result quality.
Existing out-of-place graph-based vector indexing update approaches suffer from poor update throughput due to the need to periodically merge update batches into the underlying graph index.

Building a vector data system that supports efficient and effective in-place updates is inherently challenging.
In this work, we propose \sysname{} to achieve it.
In particular, \sysname{} supports in-place graph-based vector indexing updates with consistently high update throughput and good search result quality.
The key idea of \sysname{} is \textit{decomposition facilitates consolidation}.
In particular, we introduce a vector-level update mechanism and architect \sysname{} with three core components: (i) a \tasklet{}-based execution engine, (ii) an asynchronous buffer manager, and (iii) a vector file system. 
Experimental results demonstrate that \sysname{} achieves 1.75x higher update throughput and 1.8x higher concurrent search throughput than the state-of-the-art systems on the 800M dataset, while using only 73\% of the peak memory and fewer CPU cores.




\end{abstract}

\maketitle

\pagestyle{\vldbpagestyle}
\begingroup\small\noindent\raggedright\textbf{PVLDB Reference Format:}\\
\vldbauthors. \vldbtitle. PVLDB, \vldbvolume(\vldbissue): \vldbpages, \vldbyear.\\
\href{https://doi.org/\vldbdoi}{doi:\vldbdoi}
\endgroup
\begingroup
\renewcommand\thefootnote{}\footnote{\noindent
This work is licensed under the Creative Commons BY-NC-ND 4.0 International License. Visit \url{https://creativecommons.org/licenses/by-nc-nd/4.0/} to view a copy of this license. For any use beyond those covered by this license, obtain permission by emailing \href{mailto:info@vldb.org}{info@vldb.org}. Copyright is held by the owner/author(s). Publication rights licensed to the VLDB Endowment. \\
\raggedright Proceedings of the VLDB Endowment, Vol. \vldbvolume, No. \vldbissue\ %
ISSN 2150-8097. \\
\href{https://doi.org/\vldbdoi}{doi:\vldbdoi} \\
}\addtocounter{footnote}{-1}\endgroup


\section{Introduction} \label{sec:intro}

Approximate nearest neighbor search (ANN) is widely utilized in various domains, ranging from recommendation systems~\cite{li2021embedding,xiang2019accelerating} to sparse attention in LLM inference~\cite{infllm,RA}.
In particular, deep learning models can transform multi-modal data into vectors in a space of hundreds or thousands of dimensions.
The similarity of these embedded vectors is measured by their distance.
Then, the top-$k$ query is widely used to retrieve the $k$ nearest neighbors of a given query vector. 
The crux of the top-$k$ query over a large-scale vector data set is the underlying \emph{index}, which effectively organizes the vector embeddings for efficient searching.
Among all types of ANN index, graph-based indexes are \emph{de facto} standards due to their superior query efficiency and result accuracy~\cite{wang2024starling, jayaram2019diskann, singh2021freshdiskann}.

The rapid evolution of data in many applications demands high-throughput vector updates.
For example, Alibaba's smart city application generates up to 100 million new vector records per day~\cite{wei2020analyticdb}, and the ANN systems built on email servers need to host trillion-scale vectors with real-time updates~\cite{xu2023spfresh}.
To address these challenges, various approaches have been proposed to enable vector updates on the underlying index.
In this work, we focus on enabling updates for on-disk, graph-based vector indexes, as prior studies have shown that cluster-based indexes cannot simultaneously achieve high update throughput and strong search quality~\cite{pipeann}.

The representative graph-based solutions are summarized in Table~\ref{fig:existing-solutions-overview}.
DiskANN~\cite{singh2021freshdiskann} uses an \textit{out-of-place} strategy: it accumulates insertions and deletions in memory and periodically merges them offline into the global on-disk graph index. 
While it avoids rebuilding the index from scratch, it still demands substantial memory for temporary structures and incurs costly merge operations.
To reduce this overhead, OdinANN~\cite{guo2025odinann} introduces a direct approach to insert vectors into the on-disk graph index.
However, it still employs an out-of-place approach to apply deletion updates.


\begin{table}[]
\centering
\caption{Overview of graph-based vector update approaches.}\label{fig:existing-solutions-overview}
\begin{tabular}{|c|c|c|}
\hline
Solution   & insert       & delete       \\ \hline\hline
DiskANN~\cite{singh2021freshdiskann} & out-of-place & out-of-place \\ \hline
OdinANN~\cite{guo2025odinann}    & in-place     & out-of-place \\ \hline
IP-DiskANN~\cite{xu2025place} & ---          & in-place     \\ \hline
\textbf{\sysname{} (Our work)} & \textbf{in-place}     & \textbf{in-place}     \\ \hline
\end{tabular}
\end{table}

Fundamentally, out-of-place strategies are ill-suited for large-scale real-time vector index updates.
First of all, the merge cost increases linearly with the size of the data.
For instance, in the state-of-the-art OdinANN, scaling the index from 100M to 800M vectors makes 20M updates increase merge time by 2.67$\times$, which becomes even larger than its in-place processing time.
Consequently, if the update arrival speed is faster than they can be merged, the unprocessed updates will accumulate uncontrollably.
To make matters worse, the size of the update batch affects update speed, graph quality, and memory usage.
For example, the batch size of existing approaches ranges from 0.1\%~\cite{yu2025topology} to 20\%~\cite{guo2025odinann} of the total dataset.
Thus, it is difficult for users to set a proper batch size for their usage.

Considering the above-mentioned limitations of out-of-place strategies, 
it is intuitive to devise an in-place graph update approach.
Unfortunately, it is inherently challenging to build a vector data system that supports in-place vector updates as
the graph-based vector indexes are singly-linked, deletions are difficult because identifying in-neighbors of a vertex requires a full scan.
A heuristic in-place deletion algorithmic idea was proposed by IP-DiskANN~\cite{xu2025place}.
However, it remains far from practical use, as IP-DiskANN completely ignores the handling of insertions.
A further step is to combine OdinANN’s direct insertion mechanism~\cite{guo2025odinann} with IP-DiskANN’s in-place deletion algorithm~\cite{xu2025place} to enable in-place updates of the graph index.
However, it still does not address the following four technical challenges to support efficient and effective in-place graph index updates.

\stitle{(1) Excessive Vector Visits} 
Each in-place vector insertion or deletion traverses a set of neighboring nodes in the proximity graph index.
Due to the inherent structure of proximity graph indexes, many of these nodes are repeatedly accessed across updates. 
Out-of-place update approaches address it by consolidating updates into a batch and visiting each node in the graph only once via a full scan. 
Consequently, the first challenge in enabling efficient in-place graph-based vector updates is eliminating redundant node visits.

\stitle{(2) Suboptimal CPU utilization} 
Existing graph index update solutions significantly under-utilize CPU resources—either due to I/O stalls~\cite{guo2025odinann} or contention on a global lock~\cite{ipdiskann-code2025}. 
Thus, the second challenge in building an in-place graph index update system is to fully exploit CPU resources to achieve both high update throughput and good-quality search results.

\stitle{(3) Ineffective Cache Scheme} 
Leveraging in-memory caching to improve the performance of on-disk graph-based indexes is an intuitive approach. 
However, existing methods either statically cache ``hot'' vectors without supporting dynamic updates~\cite{wang2024starling, jayaram2019diskann}, or cache vectors temporarily per query~\cite{guo2025odinann}. 
Hence, the third challenge is devising an efficient buffer manager for in-place vector updates.

\stitle{(4) Inefficient Data Layout} 
Interestingly, we observed that the data accessed from the graph index during searches differs from that accessed during updates. 
For instance, raw vector data is used only in search operations, i.e., to ensure high result quality, 
and is not involved in updates. 
Yet, all existing systems tightly couple the index topology with raw vector data. 
Then, the fourth challenge is designing an efficient data layout that enables both high-performance search and efficient in-place updates in the underlying graph index.

\begin{figure}
  \small
  \centering
  \includegraphics[width=0.85\columnwidth]{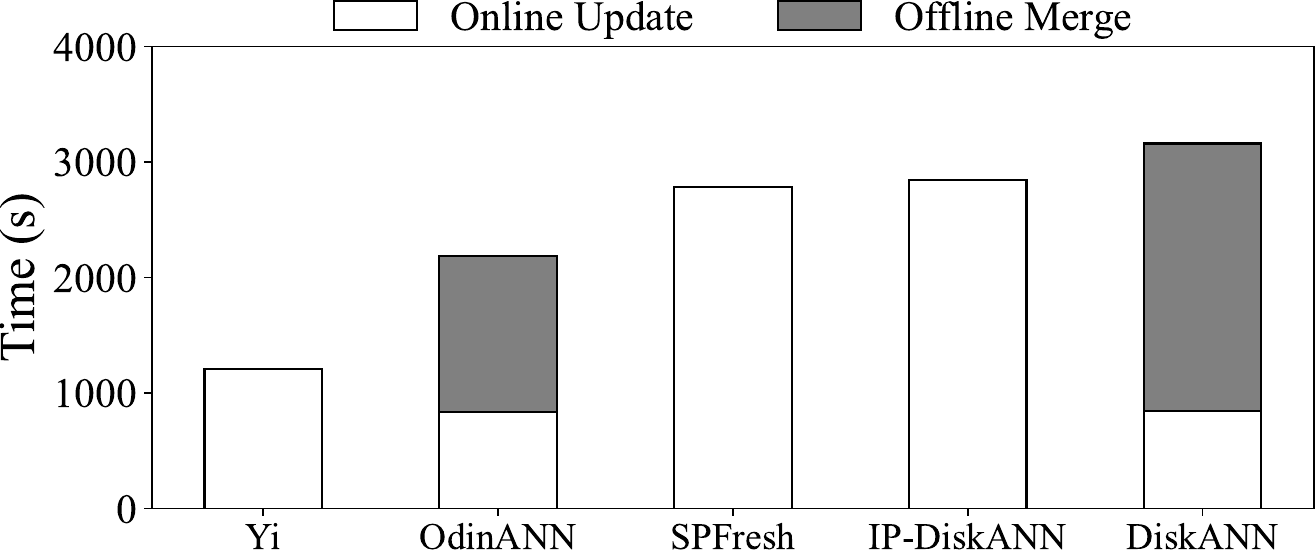} 
  \caption{Elapsed time to replace 1M data in a 100M index. All systems run with their default parameters and IP-DiskANN is implemented upon the components of our system \sysname{}.
   For the end-to-end update processing time, \sysname{} is 2.30 and 1.81 times faster than the state-of-the-art cluster-based (i.e., SPFresh) and graph-based  (i.e., OdinANN) index update solutions, respectively.
  We will introduce online update and offline merge phases in Section~\ref{sec:limitations}.}
  \label{fig:end-to-end}
  \trim 
\end{figure}

In this work, to overcome these challenges, we propose \sysname{} to support in-place graph-based vector index updates with high update throughput and good search quality.
We first conduct an in-depth analysis of the limitations of existing out-of-place solutions and the issues faced to build an in-place system.
We propose the core vector-level update mechanism by following the \textit{decomposition facilitates consolidation} principle. 
In particular, it divides query-level updates into fine-grained, independent vector-level operations.
We then implement the vector-level update mechanism into \sysname{}, which contains three major components: (1) a \tasklet{}-based execution engine, (2) an asynchronous buffer manager, and (3) a vector file system.
Finally, we conduct extensive experimental studies to demonstrate the superiority of \sysname{} over all existing solutions (see Figure~\ref{fig:end-to-end}).
Specifically, \sysname{} delivers 1.75x higher update throughput and 1.76x higher concurrent search throughput than the state-of-the-art graph index update solution OdinANN~\cite{guo2025odinann} on the 800M dataset, while using only 73\% of the peak memory and fewer CPU cores.


The rest of the paper is organized as follows. 
We discuss background and motivation of our work in Section~\ref{sec:background}, present the core idea of vector-level update mechanism in Section~\ref{sec:vector-level}, introduce our system \sysname{} in Section~\ref{sec:system}, conduct experiments in Section~\ref{sec:exp} and conclude this work in Section~\ref{sec:con}.

\section{Background and Motivation}\label{sec:background}

Given a vector dataset $\mathbf{D}$ and each vector $\mathbf{r} \in \mathbf{D}$ with $d$ dimensions,
the \emph{vector index} is constructed on $\mathbf{D}$ to process \textit{approximate nearest neighbor (ANN)} query efficiently.
The processing efficiency (i.e., throughput) and result accuracy (i.e., recall) are two fundamental performance metrics for ANN query.
In the literature, dozens of vector index solutions~\cite{datar2004locality,he2010scalable,kulis2009kernelized, mu2010non,raginsky2009locality, song2011multiple,wang2012semi,weiss2008spectral, xu2011complementary, babenko2014inverted, baranchuk2018revisiting, jegou2011searching, zhang2019grip, dong2011efficient, fu2019fast, hajebi2011fast, malkov2018efficient, toussaint1980relative, wang2012scalable, arya1998optimal, babenko2017product, beis1997shape, bentley1975multidimensional,dasgupta2008random, friedman1977algorithm, moore2000anchors, muja2014scalable, nister2006scalable, sproull1991refinements, wang2013trinary, iwasaki2016pruned, wang2012query,chen2018sptag} have been proposed to achieve good performance on both performance metrics.
In this work, we focus on the graph-based indexes as they have demonstrated the superiority on both metrics over other competitors~\cite{aumuller2020ann, azizi2025graph, gou2025symphonyqg, wang2021comprehensive}.
Unfortunately, it is challenging to support graph-based vector index update with high throughput and good quality.
In this section, we first introduce the key subroutines in graph-based index update systems in Section~\ref{sec:problem-statement},
then analyze the limitations of existing out-of-place solutions in Section~\ref{sec:limitations}, and last investigate the issues to build an efficient in-place graph-based index update system in Section~\ref{sec:challenges}.

\subsection{Key Subroutines for Index Updates}\label{sec:problem-statement}
Graph-based vector index update systems consist of two core subroutines: \textsf{Insert} and \textsf{Delete}.
In particular, the \textsf{Insert} subroutine invokes the \textsf{Search} subroutine during its processing.
We elaborate on these subroutines as follows.

\stitle{\textsf{Search} Subroutine} 
It retrieves the nearest neighbors of a given query vector by traversing the graph.
Specifically, the traversal starts from an entry point and explores neighboring vertices based on their similarity to the query vector. 
This process terminates when the stopping criterion is satisfied, 
e.g., no candidate vector node has a smaller distance than the distance to the $k$-th nearest neighbor in the result set.

\stitle{\textsf{Insert} Subroutine}
It first invokes \textsf{Search} subroutine to retrieve the nearest neighbors of the inserting vector.
Next, inverted connections are established from these retrieved neighbors to the inserted vector,
and the pruning algorithm will be invoked if the degree of any affected vectors exceeds the predefined maximum degree of the graph~\cite{jayaram2019diskann}.

\stitle{\textsf{Delete} Subroutine} 
It first marks the corresponding vector on the graph as deleted, 
then establishes the connection from the ingoing neighbors of the deleted vector to all its outgoing neighbors.
Similarly, the pruning algorithm will also be invoked if the degree of affected vectors exceeds the maximum degree during the deletion processing.


\subsection{Limitations of Out-of-place Updates}\label{sec:limitations}
\begin{figure}
  \small
  \centering
  \includegraphics[width=0.85\columnwidth]{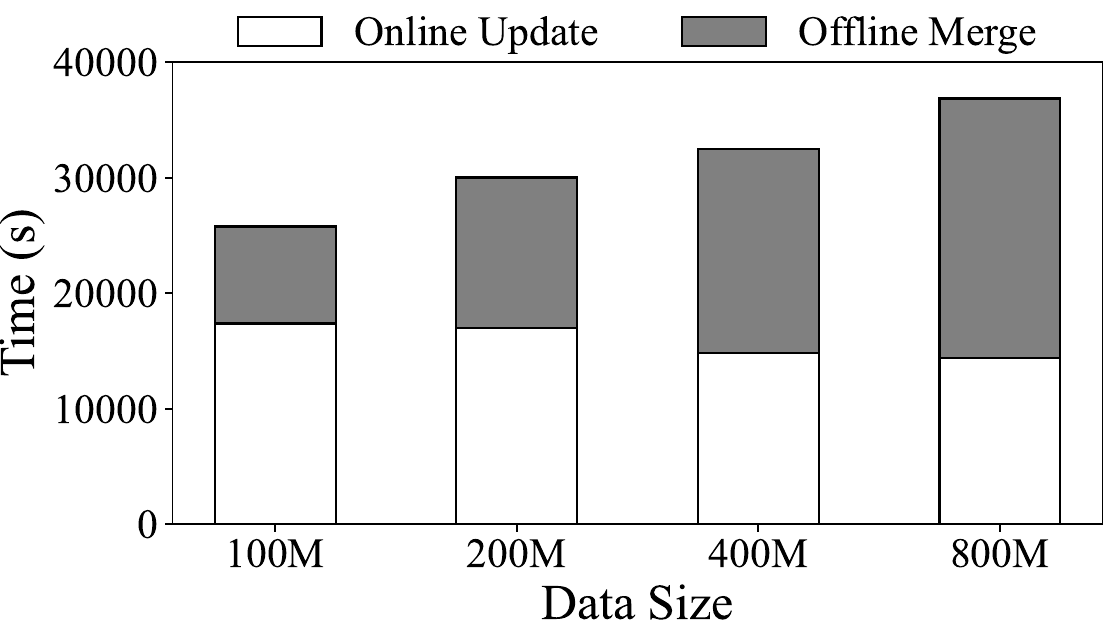} 
  \caption{Inserting and Deleting 20M vectors in OdinANN with different index size on DEEP with 10 cores.}
  \label{fig:out-of-place}
\end{figure}

The \textit{out-of-place} update methods have been widely used in existing systems~\cite{singh2021freshdiskann, guo2025odinann, yu2025topology}. 
Instead of applying \textsf{Insert} and \textsf{Delete} subroutines individually and immediately, updates are processed in batches.
Each batch of updates consists of an online update stage and an offline merge stage.
We illustrate the general idea of out-of-place updates by introducing these two stages in DiskANN~\cite{singh2021freshdiskann}.

\sstitle{Online Update Stage} Each insert adds the new vector to the in-memory temporary index, while each delete marks the deleted vectors by using a delete list. 

\sstitle{Offline Merge Stage} Once the accumulated updates exceed a predefined threshold (a.k.a., batch size), the system merges the temporary structures in memory into the on-disk index with full scans.

To improve the update efficiency of DiskANN, Greator~\cite{yu2025topology} introduces an auxiliary graph topology that speeds up updates on extremely small batches (e.g., 0.1\% of the index size) by quickly identifying the affected vectors;
OdinANN~\cite{guo2025odinann} presents a direct insertion mechanism that reduces the memory footprint of newly inserted vectors and reduces the merge time required to integrate them.
OdinANN is the state-of-the-art graph-based vector index update system.
However, there remains substantial room for improvement.
On one hand, the offline update throughput of OdinANN falls with the increasing of the vector dataset size as the cost of offline merge stage incurs more computations and I/Os when the underlying graph index is larger. 
As shown in Figure~\ref{fig:out-of-place}, when the index grows from 100M to 800M vectors, the offline merge cost of inserting and deleting 20M vectors in OdinANN increases 2.67 times.
Moreover, the offline merge cost dominates the online update cost when the indexed data size is 800M.
On the other hand, it is non-trivial to support high-rate real-time update by out-of-place solutions, as online updates and offline updates are mutually constraining.  
In particular, if updates arrive at a constant rate $T$, queueing theory~\cite{Kleinrock1975QueueingSystems} shows that the online system can only serve them when
\begin{equation}
    T \le \frac{T_o\cdot T_f}{T_o + T_f}
\end{equation}
where $T_o$ and $T_f$ are the online and offline update throughput of the system, respectively.
If the update arrival rate exceeds this threshold, the update queue will grow indefinitely.
For OdinANN running on 10 cores with a 20M batch on a 100M vector index, the online and offline throughput are 1512 QPS and 3820 QPS, respectively.
As a result, the highest update rate should be smaller than 1083 QPS.

\subsection{Issues with In-place Updates}\label{sec:challenges}
An intuitive idea to overcome the above limitations of out-of-place update methods is to apply the updates to the underlying graph index directly (a.k.a. in-place updates).
In particular, each update query is first mapped to a dedicated \textsf{Insert} or \textsf{Delete} subroutine, 
and then the corresponding subroutine is processed individually.
From the implementation, an in-place vector update system can be built upon existing studies.
For example, the \textsf{Insert} subroutine uses the direct insert method in OdinANN~\cite{guo2025odinann}.
The \textsf{Delete} subroutine utilizes the heuristic in-place delete algorithm in IP-DiskANN~\cite{xu2025place}.
In particular, the execution of \textsf{Delete} subroutine is extremely time-consuming as it incurs a full scan.
IP-DiskANN~\cite{xu2025place} proposes a heuristic in-place delete algorithm to reduce the cost, i.e., it first identifies a subset of in-neighbors for the vectors to be deleted through a \textsf{Search} subroutine and then establishes connections and performs potential pruning only within this subset.
However, it is still far from achieving a high-throughput in-place graph-based vector index update system due to the following issues.




\stitle{{Issue 1: Excessive Vector Visits}} 
In-place updates visit significantly more vectors than out-of-place updates, as many nodes will be repeatedly visited by the processing of different subroutines. For example, using default parameters, deleting 20M vectors from the 100M DEEP dataset with IP-DiskANN visits about 4B vectors during search. 
In contrast, updating them with an out-of-place full scan visits only 100M vectors.

\stitle{{Issue 2: Suboptimal CPU Utilization}} 
According to our internal experiments, the direct insert method in OdinANN achieves only 61\% average CPU utilization when inserting 20M vectors into the 100M DEEP index, which is caused by waiting for I/O.
In addition, the in-place delete algorithm in IP-DiskANN~\cite{ipdiskann-code2025} relies on a global lock, further constraining concurrency and preventing efficient parallel updates.


\stitle{{Issue 3: Ineffective Cache Scheme}} 
Many graph-based systems~\cite{jayaram2019diskann, wang2024starling} cache frequently visited vectors when processing \textsf{Search} subroutine. However, they do not support dynamic updates. 
OdinANN~\cite{guo2025odinann} dynamically caches vectors during the search phase of each insert but discards them after each query, leaving substantial cache space to be explored.

\begin{figure}
  \small
  \centering
  \begin{tabular}{cc}
  \includegraphics[width=0.46\columnwidth]{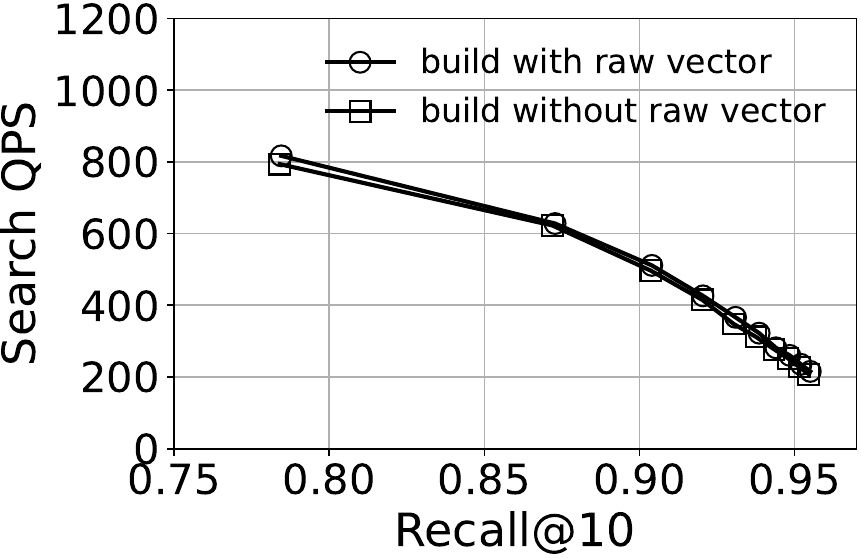}
  &
  \includegraphics[width=0.46\columnwidth]{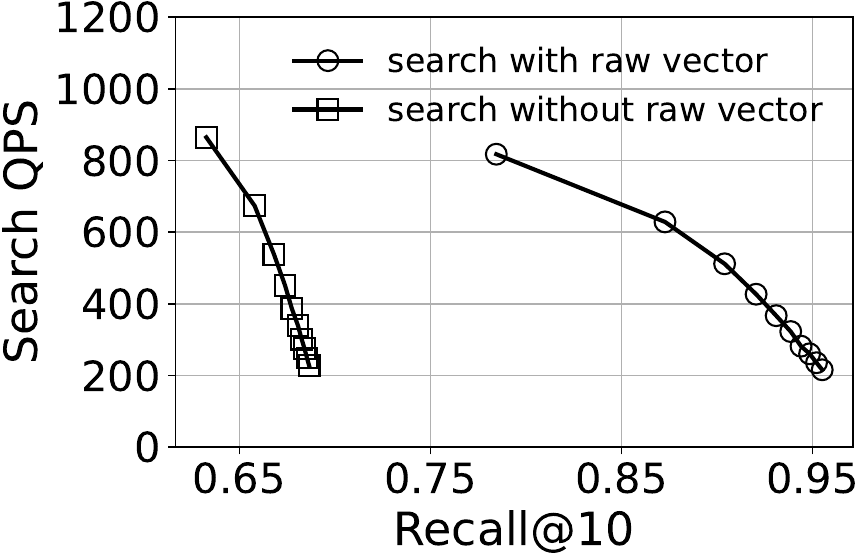}
  \\
  (a) build (update from scratch)   & (b) search \\
  \end{tabular}
  \trim 
  \caption{Effect of raw data on build and search}
  \label{fig:filesys}
  \trim 
\end{figure}

\stitle{{Issue 4: Inefficient Data Layout}}
An interesting observation from our experiments is that the necessary data accessed by search and update queries are different.
As depicted in Figure~\ref{fig:filesys}, the raw vectors are unnecessary for updates (see Figure~\ref{fig:filesys}(a)), but they significantly affect search quality (see Figure~\ref{fig:filesys}(b)).
In existing graph-based indexes, accessing a vector requires reading its corresponding block, where each entry stores both out-neighbors and raw data for the vector.
This motivates us to devise an efficient data layout to improve block access efficiency for different subroutines.

\section{Core Idea: Vector-level Update Mechanism}\label{sec:vector-level}

Logically, the straightforward in-place update solution in Section~\ref{sec:challenges} utilizes a query-level update mechanism as it processes each insert or delete query individually. 
Based on the above discussion, directly applying query-level updates is inefficient. 
In this work, we resolve these mentioned issues by devising a novel vector-level update mechanism, which follows the ``decomposition simplifies consolidation'' design principle.
In particular, prior work has shown that dividing complex operations into smaller, independent tasks improves scheduling flexibility~\cite{liu2024tao, ruan2023nu, leis2014morsel}. 
We apply this idea and treat each vector as the fundamental unit of a query to build an efficient in-place graph-based vector index update system. 
We next illustrate an insightful observation, which sheds light on us to propose the novel vector-level update mechanism.

\begin{figure*}
  \small
  \centering
  \begin{tabular}{cc}
  \includegraphics[width=\columnwidth]{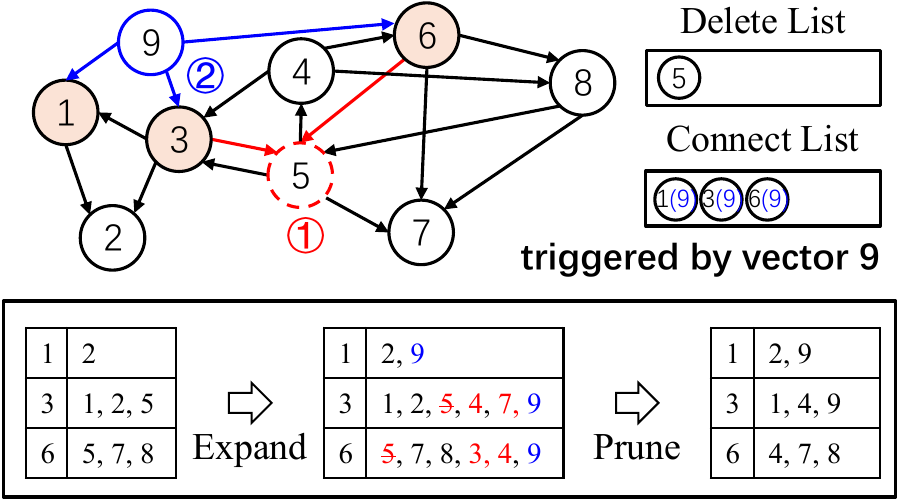}
  &
  \includegraphics[width=\columnwidth]{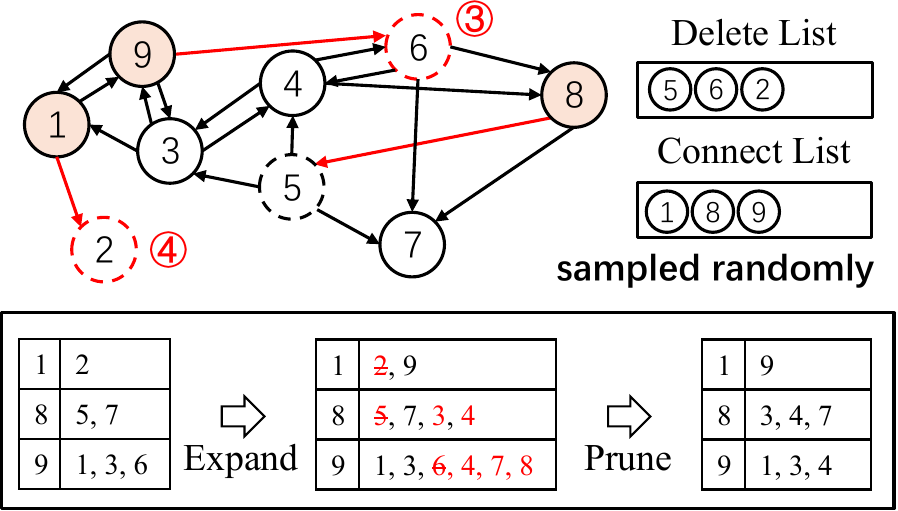}
  \\
  (a) Vector visiting triggered by inserting  & (b) Vector visiting triggered by sampling
  \end{tabular}
  \trim \trim
  \caption{Illustration of the vector-level update mechanism}
  \trim \trim
  \label{fig:abstraction}
\end{figure*}



Returning to the \textsf{Insert} and \textsf{Delete} subroutines in Section~\ref{sec:problem-statement},
we observed that the connection establishment procedures of both subroutines share a similar logic, which can be summarized as the following two steps:

\sstitle{(1) Expand} It first identifies an ANN set for a given vector 
and then expands the outgoing neighbor list for every vector in this ANN set.
For the \textsf{Insert} subroutine, the newly inserted vector will be included in the outgoing neighbor list of every vector in the ANN set.
For the \textsf{Delete} subroutine, the outgoing neighbor list of every vector in the ANN set first excludes the deleted vector, and then appends the outgoing neighbor list of the deleted vector into it.

\sstitle{(2) Prune}  For each vector in the ANN set, it applies a pruning algorithm~\cite{jayaram2019diskann} if its expanded outgoing neighbor list size exceeds the predefined maximum out-degree.

Motivated by the above key observation, we propose a vector-level update mechanism, which kills two birds (\textsf{Insert} and \textsf{Delete} updates) with one stone (a unified task). 
In particular, we unify the connection-establishing task for each vector in the ANN set of both \textsf{Insert} and \textsf{Delete} subroutines, which applies \textsf{Expand} and \textsf{Prune} steps on its out-neighbor list.

We use the example in Figure~\ref{fig:abstraction} to illustrate the above vector-level update mechanism.
To process in-place vector updates, we maintain two lists: a delete list and a connect list.
As shown in Figure~\ref{fig:abstraction}(a), we first delete vector 5 (\circledcolor{red}{red}{1}), then insert vector 9 (\circledcolor{blue}{blue}{2}) in the graph index.
To process these updates, the deleted vectors are appended to the delete list first.
For example, when vector 5 is deleted (\circledcolor{red}{red}{1}), it is not removed from storage immediately, but it is appended to the delete list. 
For the insert vector 9, its nearest neighbors (i.e., vectors 1, 3, and 6) are retrieved and appended to the connect list.
We next execute the unified connection task for every vector in the connect list to process the above \textsf{delete} and \textsf{insert} updates.
Taking vector 3 as an example,  its \textsf{connect} task first accesses its outgoing neighbor list, which is \{1,2,5\}.
It then includes the newly inserted vector 9 (i.e., \{1,2,5,9\}), which is the \textsf{expand} step of \textsf{Insert} subroutine.
It next excludes the deleted vector 5 and appends the outgoing neighbor list of vector 5 (i.e., \{3,4,7\}) into it and returns \{1,2,4,7,9\}.
Since the size of vector 3's outgoing neighbor list exceeds the pre-defined maximum out-degree 3, it then applies the pruning algorithm and returns \{1,4,9\} as the result of vector 3's \textsf{connect} task.
Similarly, the outgoing neighbor lists of vectors 1 and 6 in the connect list will be updated by their corresponding \textsf{connect} tasks, see the updated lists in Figure~\ref{fig:abstraction}(a).

An extreme case is that the connect list has too few (or even does not have) vectors, which occurs when the workload is delete-intensive or delete-only workloads.
\sysname{} randomly samples a small range of consecutive on-disk vectors and includes them in the connect list to mitigate this issue.
For example, in Figure~\ref{fig:abstraction}(b), vectors 2 (\circledcolor{red}{red}{3}) and 6 (\circledcolor{red}{red}{4}) have been deleted, but they do not have \textsf{insert} updates. 
\sysname{} then randomly selects three vectors (i.e., vectors 1, 8, 9) and includes them in the connect list, then applies the \textsf{connect} task on each of them.

Specifically, a vector is evicted from the connect list immediately after its connect task executes. 
In contrast, vectors are retained in the delete list, which is a fixed-size Least Recently Used (LRU) cache.
In particular, a deleted vector can be removed from disk immediately once it is evicted from the LRU cache. 
The rationale is that keeping a deleted vector accessible for long enough makes it almost certain that all of its in-neighbors will eventually be visited.

\stitle{Effectiveness Analysis} With the vector-level update mechanism,  we consolidate the expand and prune operations from different queries on a vector using a specific \textsf{connect} task.
It is reasonable because the proximity graph only needs vector-level snapshot isolation~\cite{guo2025odinann}.
As a result, many redundant operations are eliminated, which mitigates Issue 1 in Section~\ref{sec:challenges}.
Moreover, this mechanism serves as the foundation for our system-level design to address Issues 2, 3, and 4.



\section{Design and Implementation}\label{sec:system}

In this section, we design \sysname{} to provide system-level support and resolve the remaining issues outlined in Section~\ref{sec:challenges}.
We first provide the overall architecture of \sysname{} in Section~\ref{sec:sys-overall}, then describe three major components in it, i.e., \tasklet{}-based execution engine, asynchronous  buffer manager and vector file system in Section~\ref{sec:tasklet},~\ref{sec:buffer} and~\ref{sec:vectorfs}, respectively.

\subsection{System Overview}\label{sec:sys-overall}

\begin{figure}[t]
  \small
  \centering
  \includegraphics[width=\columnwidth]{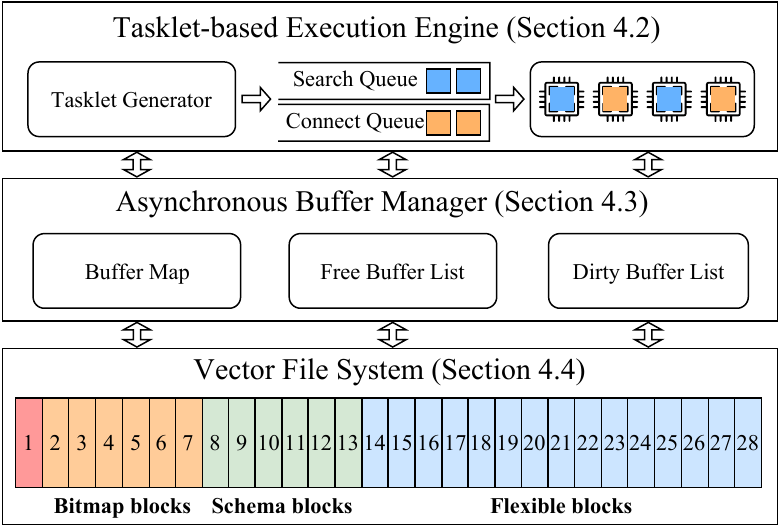}
  \trim \trim
  \caption{System architecture of \sysname{}}
  \label{fig:architecture}
  \trim \trim
\end{figure}

Figure~\ref{fig:architecture} depicts the overview of \sysname{},
which includes three key components to overcome the limitations of existing solutions.


\begin{figure*}[t]
  \small
  \centering
  \includegraphics[width=2.1\columnwidth]{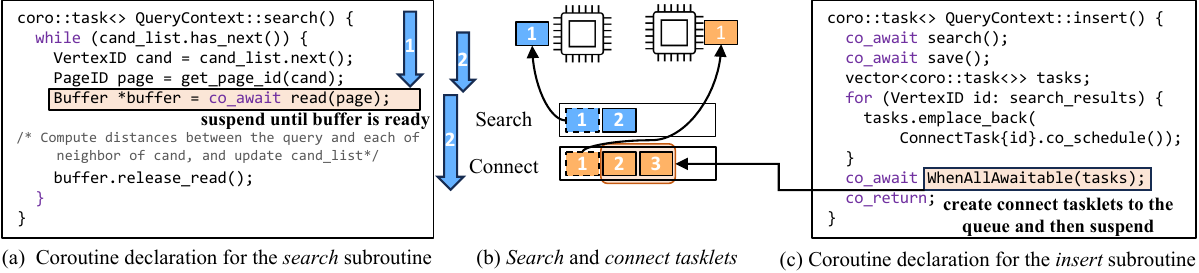}
  \trim \trim 
  \caption{\Tasklet{} generation in \sysname{}}
  \label{fig:generation}
  \trim \trim
\end{figure*}

\squishlist
\item \stitle{\Tasklet{}-based Execution Engine (Section~\ref{sec:tasklet})} 
To improve parallelism and resource utilization (Issue 2) for vector-level update tasks proposed in Section~\ref{sec:vector-level}, we introduce lightweight \tasklets{} as the basic execution units and build a \tasklet{}-based execution engine for them. 

\item \stitle{Asynchronous Buffer Manager (Section~\ref{sec:buffer})} 
To address the ineffective cache schemes in existing graph-based systems (Issue 3), we design an asynchronous buffer manager that caches vectors and pages across update operations. 
It abstracts page access protocols for vector updates and handles concurrent accesses in a lightweight, asynchronous manner.


\item \stitle{Vector File System (Section~\ref{sec:vectorfs})} 
To improve storage efficiency for vector updates (Issue 4), we design a vector file system tailored to the access patterns of graph-based vector indexes. 
In particular, it separates raw data from the navigation graph into different pages as B$^+$-tree~\cite{silberschatz2011database} and Linux inode designs.
\squishend


\subsection{\Tasklet{}-based Execution Engine}\label{sec:tasklet}

Returning to Issue~2 in Section~\ref{sec:challenges}, our goal is to design an efficient execution paradigm that improves CPU utilization.
The main difficulty is that computation in an on-disk graph-based index often wait for page I/O or page conflicts.
Prior systems have shown that search path can be decomposed into I/O issuing and frontier exploration stages to overlap I/O and computation~\cite{pipeann, guo2025odinann}.
However, efficient scheduling of the vector-level neighbor updates proposed in Section~\ref{sec:vector-level} remain an open challenge, as each update can involve variable computation (e.g., handling different numbers of deleted neighbors) and they may access pages being used by other updates.


Following the design principle of ``decomposition simplifies scheduling''~\cite{ruan2023nu, liu2024tao}, this challenge drives us to decouple the original queries into vector-level execution units to enable fine-grained scheduling.
We use \textbf{\tasklets{}} as the basic execution units of \sysname{}, following the naming convention in Tao~\cite{liu2024tao}. 
In particular, a \tasklet{} is a fine-grained piece of a query that can be scheduled independently and can suspend when waiting for page I/O or page-access conflicts. 
To realize these properties, we implement tasklets using C++20 coroutines, driven by the following reasons.

\squishlist

\item \sstitle{Easy to Use}  
We can easily convert a query into vector-level execution units by injecting \textsf{co\_await} within the coroutine declaration.

 \item \sstitle{Flexible} Once a coroutine is suspended, its underlying thread can execute other coroutines.
 This flexibility simplifies the concurrency control that will be introduced in Section~\ref{sec:buffer}.
 
\item \sstitle{Lightweight} The operations of suspending, resuming, and spawning coroutines introduce merely a few nanoseconds of overhead~\cite{he2020corobase, jonathan2018exploiting}.

\squishend

Specifically, we decompose the original \textsf{search} and \textsf{insert} queries into fine-grained \textsf{search} and \textsf{connect} \textbf{\tasklets{}} using C++20 coroutines. Figure~\ref{fig:generation} shows how these \tasklets{} are generated.
Next, we describe \textsf{search} and \textsf{connect} \tasklets{} in detail.


\sstitle{\textsf{Search} \tasklets{}} 
Figure~\ref{fig:generation}(a) shows the coroutine definition \textsf{search()}.
Its execution logic is the same as the beam search proposed in DiskANN~\cite{jayaram2019diskann} (assume that the beam width is 1).
In particular, this coroutine gradually explores the graph, checking nearby vertices, and adding promising ones to a candidate list (\textsf{cand\_list}) until the candidate list converges.

During each round of the search, the coroutine calls \textsf{read()} using \textsf{co\_await}.
This suspends the coroutine so that the system can handle I/O or resolve read/write conflicts (see Section~\ref{sec:buffer}).
We refer to the piece of code executed between two such suspensions as a \textsf{search} \tasklet{}.

In Figure~\ref{fig:generation}(a), \tasklet{} 1 covers the execution from the start of the coroutine to the first \textsf{co\_await}.
\tasklet{} 2 is the segment executed from that \textsf{co\_await} to the suspension in the next loop iteration.
Whenever a \tasklet{} is ready to continue (i.e., it is I/O ready and conflict-free), it is pushed back into the \tasklet{} queue, where CPUs can pick it up and resume its execution (see Figure~\ref{fig:generation}(b)).



\sstitle{\textbf{Connect} \tasklets{}} Figure~\ref{fig:generation}(c) shows the coroutine definition of \textsf{insert()}.
In particular, it first invokes \textsf{search()} coroutine via \textsf{co\_await}, which generates \textsf{search} \tasklets{} as discussed before.
After storing the vector into the storage via the coroutine \textsf{save()}, it generates multiple \textsf{connect} \tasklets{} as highlighted in Figure~\ref{fig:generation}(c).
Each \textsf{connect} \tasklet{} is responsible for performing the prune operation for one of the vectors in the connect list in Figure~\ref{fig:abstraction}.





\stitle{Effectiveness Analysis} With this mechanism, both \textsf{search} and \textsf{connect} \tasklets{} operate at the vector level.
Specifically, suspension enables the system to hide the I/O waiting time, while \textsf{connect} \tasklets{} for different vectors can be executed in parallel.
This improves CPU utilization by fine-grained scheduling, which addresses Issue~2 in Section~\ref{sec:challenges}.


\begin{figure*}[t]
  \small
  \centering
  \includegraphics[width=2.0\columnwidth]{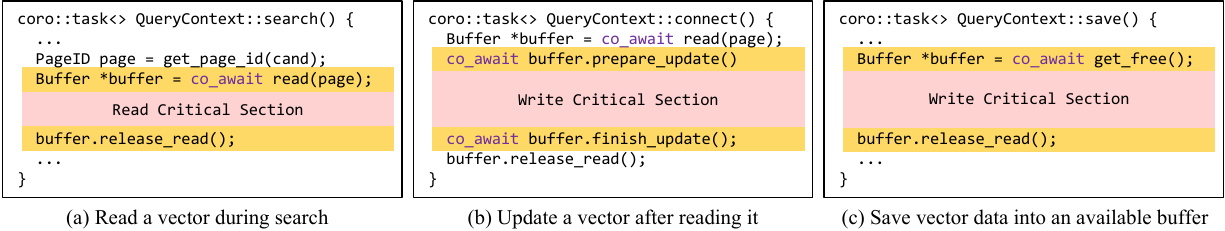}
  \trim \trim 
  \caption{Page access protocol in \sysname{}}
  \label{fig:page-protocol}
  \trim \trim 
\end{figure*}

\subsection{Asynchronous Buffer Manager}\label{sec:buffer}

Returning to Issue~3 in Section~\ref{sec:challenges}, we want a dynamic buffer that can efficiently persist hot pages across queries for reuse.

A straightforward idea is to adopt a database-like design that caches pages under a fixed memory budget.
However, buffer pools in traditional database systems are ill-suited for proximity graphs because their transactional semantics differ~\cite{guo2025odinann}.
In particular, the vector-level update mechanism proposed in Section~\ref{sec:vector-level} does not require each operation to observe one globally consistent view of the index.
Moreover, the synchronized cache mechanism used in OdinANN~\cite{guo2025odinann} is not efficient as discussed in Section~\ref{sec:challenges} (see  Issue~2), and its synchronized design will degenerate the execution of \tasklets{} into threads~\cite{cppcoroutine}.

To develop an efficient caching solution for our vector-level update mechanism, we first abstract protocols based on the semantic requirements of vector updates and our tasklet execution engine in Section~\ref{sec:protocal-abstraction}, then introduce our design that ensures these semantics in Section~\ref{sec:protocal-design}.

\trim 
\subsubsection{Protocol Abstraction}\label{sec:protocal-abstraction}

In this section, we propose several protocols  to simplify execution breakdown logic of \tasklets{}.
We identify three representative page access patterns and propose three corresponding protocols, as illustrated in Figure~\ref{fig:page-protocol}.


\squishlist
\item Each \textsf{search} \tasklet{} reads a vector in each loop and calculates distances.
We protect this ``read critical section'' as depicted in Figure~\ref{fig:page-protocol}(a).

\item Each \textsf{connect} \tasklet{} expands and prunes the neighbors of a vector,  then modifies its neighbor list.
We protect this ``write critical section'' as depicted in Figure~\ref{fig:page-protocol}(b).

\item Each \textsf{save()} coroutine in \textsf{insert} allocates a new page when the free page pool is empty, and this ``write critical section'' is protected (Figure~\ref{fig:page-protocol}(c)).
\squishend

It is not trivial to achieve the above protocols. 
In particular, we summarize the page-accessing behaviors below:
\begin{itemize}[itemsep=0pt, parsep=0pt, topsep=0pt, partopsep=0pt]
    \item[\circled{1}] Read a page from disk into a buffer.
    \item[\circled{2}] Read the page content from a buffer.
    \item[\circled{3}] Modify the page content in a buffer.
    \item[\circled{4}] Write a buffer to disk.
    \item[\circled{5}] Get a page corresponding to a free page on disk.
    \item[\circled{6}] Evict a buffer that is not in use.
\end{itemize} 

Based on these behaviors, there are two major challenges.

\stitle{Challenge 1: Avoiding Aggressive Eviction} 
Graph updates often reuse pages near the entry point~\cite{jayaram2019diskann}. 
If these idle hotspot pages are evicted too soon~\cite{guo2025odinann}, it triggers repeated SSD reads and reduces update throughput.

\begin{table}[]
\centering
\caption{Conflict analysis between pairs of operations}\label{tab:conflict}
\begin{tabular}{|c|c|c|c|c|}
\hline
    & \circled{1} & \circled{2} & \circled{3} & \circled{4}   \\ \hline
\circled{1} & WW (M)  & RW (M)  & WW (M)  & RW (M)    \\ \hline
\circled{2} &     & \checkmark  & RW (M)  & \checkmark    \\ \hline
\circled{3} &     &     & WW (M)  & RW (M)    \\ \hline
\circled{4} &     &     &     & WW (D) \\ \hline
\end{tabular}
\end{table}

\stitle{Challenge 2: Non-Blocking Conflict Resolution} 
Table~\ref{tab:conflict} shows the conflict matrix, where $\checkmark$ indicates no conflict, RW represents read-write conflicts, WW represents write-write conflicts, and (M) and (D) denote conflicts occurring in memory and on disk, respectively.
Although traditional locks can protect these page accesses, they block worker threads during page I/O or modification, which conflicts with our \tasklet{}-based design.

\subsubsection{Achieving Protocol Semantics}\label{sec:protocal-design}

To achieve the semantics of the protocols and address Challenges 1 and 2 in Section~\ref{sec:protocal-abstraction}, we propose an asynchronous buffer manager.


\begin{figure}[t]
  \small
  \centering
  \includegraphics[width=0.9\columnwidth]{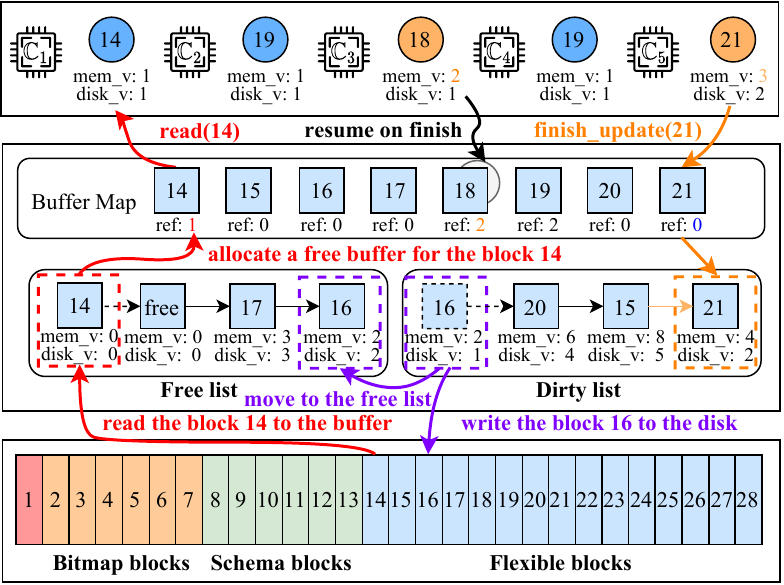}
  \trim \trim
  \caption{Overview of the buffer manager in \sysname{}.}
  \label{fig:buffer-overview}
  \trim \trim 
\end{figure}

\stitle{Overview} Figure~\ref{fig:buffer-overview} illustrates the buffer manager in \sysname{}.
It maintains a static number of fixed-size buffers (8 in this example) and has three major components: a buffer map, a free list, and a dirty list.
In particular, the buffer map traces the buffer address, the free list stores the buffers that can be directly allocated and the  dirty list stores the buffers that are modified but have not been written to disk.
Each buffer stores the content of a page.
For simplicity, we denote the buffer holding page 18 as buffer 18.
Next, we discuss the buffer eviction mechanism in our buffer manager.

\stitle{Buffer Eviction} As shown in Figure~\ref{fig:buffer-overview}, each buffer maintains a \textsf{ref} counter that records how many \tasklets{} are using it or intend to use it.
For example, it is 2 for buffer 18 because the CPU $\mathbb{C}_2$ is modifying it and another \tasklet{} is suspended on it.
Specifically, a buffer is added to the dirty list or the free list only when its \textsf{ref} drops to zero, and only buffers on the free list are eligible for eviction.

The above mechanism offers several benefits.
First, the buffers that are being accessed are never evicted, ensuring that the behaviors \circled{5} and \circled{6} are safe.
Second, it supports a predefined fixed-size memory budget, and even if a buffer is not currently referenced by any \tasklets{}, it can still remain in memory. 
This prevents hot pages (e.g., those near the entry point) from being evicted simply because they are temporarily not accessed, as in OdinANN~\cite{guo2025odinann}.
Thus, it effectively addresses Challenge 1 in Section~\ref{sec:protocal-abstraction}.

\begin{figure}[t]
  \small
  \centering
  \includegraphics[width=0.8\columnwidth]{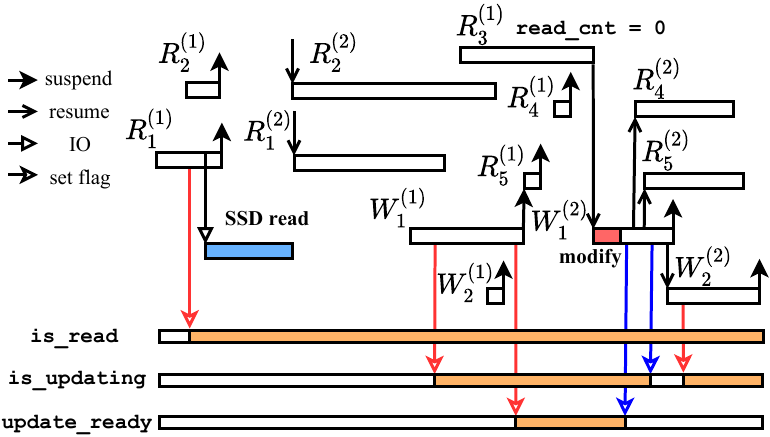}
  \trim \trim 
  \caption{Concurrency control of a specific buffer $B$}
  \trim \trim 
  \label{fig:conflict}
\end{figure}

\stitle{Concurrency Control} 
Figure~\ref{fig:conflict} shows how conflicts on a single buffer $B$ are resolved.
Here, $W$ and $R$ denote the write and read phases of the \textsf{search} and \textsf{insert} queries, and superscripts distinguish different \tasklets{} (each suspension or resumption produces a new one).
At a high level, our strategy is to \textit{suspend on conflict}, i.e., a \tasklet{} resumes only when its data is in memory and conflict-free.
In particular, suspensions are achieved by \textsf{co\_await}ing our page access protocols.
Next, we describe our protocol design in Figure~\ref{fig:page-protocol} to explain how they interact to manage page access and resolve conflicts.

\stitle{\textsf{read()}} It returns the corresponding page by a given identifier.
After \textsf{read()} returns the page, the page content can be safely accessed until \textsf{release\_read()} is called; the \textsf{read\_cnt} flag is incremented on return and decremented on release.
The call may suspend if the page is not present in the cache (see $R_1^{(1)}$ and $R_2^{(1)}$) or if it is currently being modified or is scheduled to be modified by others (see $R_4^{(1)}$ and $R_5^{(1)}$).
In particular, each suspension generates a new \tasklet{} (e.g., $R_1^{(1)}$ generates $R_1^{(2)}$).
Specifically, the conflict \circledcolor{blue}{blue}{1}-\circledcolor{blue}{blue}{1} is prevented by employing a compare-and-swap (CAS) operation to set the \textsf{is\_read} flag, ensuring that the disk read request is issued only once.
For example, only $R_1^{(1)}$ successfully sets the flag and issues the SSD read.
In addition, because $R_1$ and $R_2$ remain suspended until the read completes, conflicts \circledcolor{blue}{blue}{1}-\circledcolor{blue}{blue}{2} and \circledcolor{blue}{blue}{1}-\circledcolor{blue}{blue}{3} are avoided.

\stitle{\textsf{release\_read()}}
When invoked, both \textsf{read\_cnt} and \textsf{ref} are decremented.
If \textsf{read\_cnt} drops to 0 and \textsf{update\_ready} is set, the suspended \tasklet{} is resumed to perform the update.
Furthermore, once \textsf{ref} reaches 0, the buffer is inserted into either the dirty list or the free list.
For example, buffer 21 is placed into the dirty list, because its memory version is 4, while its disk version is 2.

\begin{figure*}[t]
  \small
  \centering
  \includegraphics[width=1.7\columnwidth]{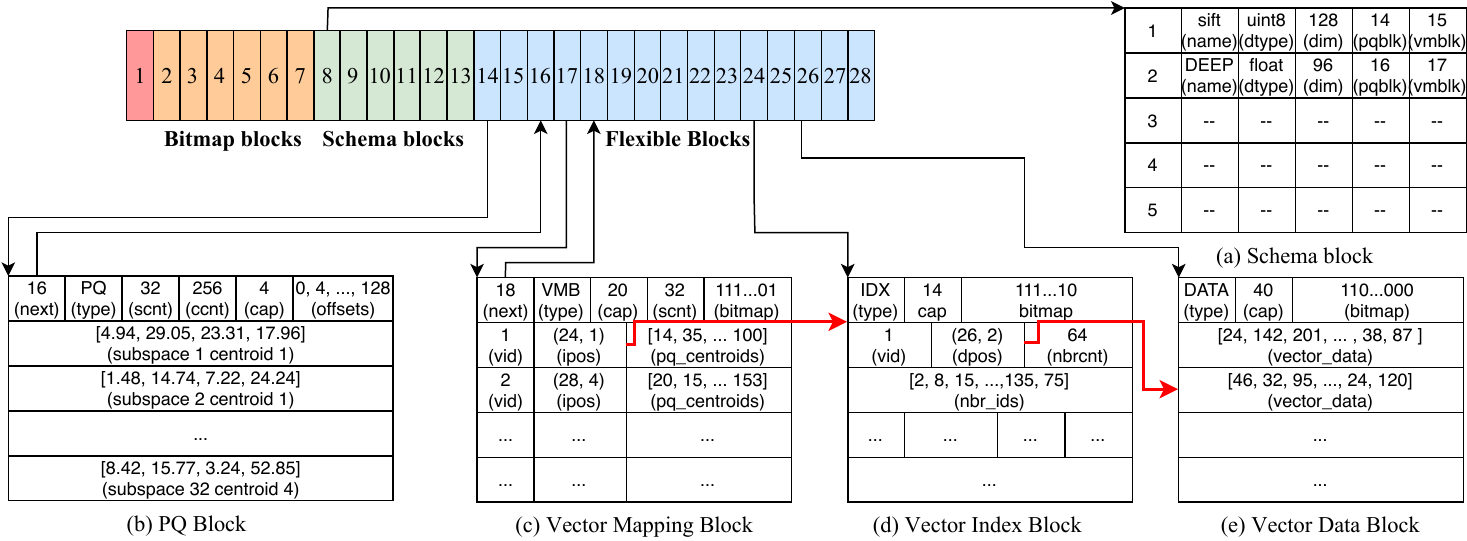}
  \trim \trim 
  \caption{Illustration of vector file system}
  \label{fig:vectorfs}
  \trim \trim
\end{figure*}

\stitle{\textsf{prepare\_update()}}
It is invoked when we want to modify a buffer content, as depicted in Figure~\ref{fig:page-protocol}.
To avoid the conflict \circledcolor{blue}{blue}{3}-\circledcolor{blue}{blue}{3}, we use a CAS operation to set a flag \textsf{is\_updating} to ensure that only one task can be scheduled to modify the page at a time.
For example, $W_1^{(1)}$ successfully sets the flag, while $W_2^{(1)}$ fails to set it and suspends itself when calling \textsf{prepare\_update()}.
The vector update operation is safe between \textsf{prepare\_update()} and \textsf{finish\_update()}.
In particular, we adopt a copy-on-write (COW) mechanism, where updates are first written to a temporary copy and only applied to the original page once the update is confirmed, as we discuss next.

\stitle{\textsf{finish\_update()}}
It is invoked to signal that the update has been confirmed and the modified page copy is ready to be committed to the original page.
However, the commit must be deferred because the page may still be under active reads (e.g., $R_{3}^{(1)}$), and further read operations may arrive concurrently (e.g., $R_{4}^{(1)}$ and $R_{5}^{(1)}$).
To handle this, we set the \textsf{update\_ready} flag, which causes subsequent readers to suspend (e.g., $R_{4}^{(1)}$ and $R_{5}^{(1)}$).
Once the last active reader (i.e., $R_{3}^{(1)}$) finishes, it applies the update on the original page content.
After such modification, other \tasklets{} are resumed (i.e., $R_4^{(2)}$, $R_5^{(2)}$, and $W_2^{(2)}$).
This mechanism avoids the \circledcolor{blue}{blue}{2}-\circledcolor{blue}{blue}{3} conflict by suspending all other \tasklets{} during the update of the original page.

\stitle{\textsf{get\_free()}} 
When inserting a vector and storing it via \textsf{save()}, as illustrated in Figure~\ref{fig:page-protocol}(c), a new page may be required. In such cases, it is acquired by invoking the \textsf{get\_free()} coroutine.
In particular, it gets an available buffer from the free list to save its content and increase the counters \textsf{ref} and \textsf{read\_cnt} by 1, respectively.

Besides the behaviors in the above protocol interfaces, buffer writing (the behavior \circled{4}) will be done asynchronously.
In particular, we generate auxiliary coroutines to write the buffers from the dirty list in head-to-tail order.
As shown in  Figure~\ref{fig:buffer-overview}, page 16 is written to the disk.
Subsequently, the disk version of the buffer is synchronized with its memory version, after which the buffer is added to the free list.
In particular, disk write is another type of memory read, which can be protected by \textsf{read()} and \textsf{release\_read()}.
In this way, it prevents the conflict \circledcolor{blue}{blue}{1}-\circledcolor{blue}{blue}{4} as \circled{1}-\circled{2}, \circledcolor{blue}{blue}{3}-\circledcolor{blue}{blue}{4} as \circled{3}-\circled{2}.
In addition, \circledcolor{blue}{blue}{4}-\circledcolor{blue}{blue}{4} will not happen because a buffer that is being written is never placed in the dirty list.

Based on the above mechanisms, we resolve all pairwise conflicts in Table~\ref{tab:conflict}, as the {\color{blue} blue circle pairs} presented above.
Thus, we address Challenge 2 in Section~\ref{sec:protocal-abstraction}.

\stitle{Effectiveness Analysis} With our asynchronous buffer manager, the page can be effectively cached in memory within a fixed memory budget, and support dynamic vector updates, which resolves Issue 3 in Section~\ref{sec:challenges}.
In addition, write requests from dirty list can be dispatched asynchronously in parallel to exploit SSD parallelism, and these I/Os are fully overlapped with computation, which in turn alleviates Issue 2 in Section~\ref{sec:challenges}.

\subsection{Vector File System}\label{sec:vectorfs}

Returning to Issue 4 in Section~\ref{sec:challenges}, we need to reconsider the vector layout for efficient updates.
We first present the layout design in Section~\ref{sec:fs-layout}, then introduce the runtime behaviors to access it in Section~\ref{sec:fs-runtime}.
In particular, we use the term \textit{block} to refer to on-disk pages in this section.

We use the term \emph{vector file system} to describe a logical storage layer.
It provides file-system-like functions for vector indexes, including metadata management, block allocation and recycling, slot-level space tracking, and logical-to-physical address translation.
These responsibilities resemble the core functions of a file system, but they are specialized for vector-index storage.

\subsubsection{Layout Design}\label{sec:fs-layout}
The layout design is motivated by the observation that, during vector updates, traversing the graph using only in-memory compressed data is sufficient, as illustrated in Figure~\ref{fig:filesys}. 
However, existing systems~\cite{jayaram2019diskann, wang2024starling, xu2025place, yu2025topology} typically store raw vectors together with the graph topology.
To address this, our design separates navigation data (i.e., graph topology) from storage data (i.e., raw vectors), placing them in distinct pages.
This layout also echoes the design principles of B+-trees~\cite{silberschatz2011database} and Linux inodes. 
Moreover, because vector records are fixed-size, we use a bitmap-based position tracking method to manage slot usage for vector placement.
Specifically, we follow a file system–style design.
Figure~\ref{fig:vectorfs} shows the layout of our vector file system.
In particular, the blocks are organized into four categories:

\sstitle{Meta Block} It is the first block storing the meta information such as the block size, the starting positions and sizes of bitmap blocks and schema blocks, etc.

\sstitle{Bitmap Blocks} 
The number of bitmap blocks is based on the total disk size. 
Each bit represents the occupancy status of a flexible block, which will be explained shortly. 
For example, the first bit in block 2 indicates whether block 14 is available for allocation.

\sstitle{Schema Blocks} 
Each schema stores the index description, as shown in Figure~\ref{fig:vectorfs}(a).
In the vector file system, each index is treated as a \emph{file}, and each schema entry holds the metadata describing a vector file, which is treated as the ``inode'' used to traverse the graph when accessing a vector.

\sstitle{Flexible Blocks} These blocks are used to store various types of data within a vector file. Specifically, they can be categorized into four distinct types, as outlined below.

\sstitle{PQ Blocks} Figure~\ref{fig:vectorfs}(a) illustrates the structure of a PQ block. The blocks are organized in a linear chain, and each block stores the ID of its subsequent block, as indicated by the \textsf{next} field. 
Additionally, each block contains metadata—such as type, subspace count (\textsf{scnt}), centroids per subspace (\textsf{ccnt}), capacity (\textsf{cap}), and subspace offsets—followed by the centroids stored sequentially.

\sstitle{Vector Mapping Blocks} 
Figure~\ref{fig:vectorfs}(b) illustrates the layout of a vector mapping block, which is also organized as a linear chain. Each block contains fixed-size slots and a bitmap indicating slot availability. 
Each slot stores the vector ID (\textsf{vid}), index position (\textsf{ipos}), and compressed PQ subspace IDs (\textsf{pq\_centroids}). 
In particular, the index position is represented by a tuple in terms of \textsf{(PageID, offset)}.
For instance, (24, 1) refers to the first slot in block 24.

\sstitle{Vector Index Blocks} 
Figure~\ref{fig:vectorfs}(c) shows the layout of a vector index block, where graph-based indexes are stored in fixed-size slots. Each slot contains the vector ID (\textsf{vid}), data position (\textsf{dpos}), number of neighbors (\textsf{nbrcnt}), and the IDs of outgoing neighbors (\textsf{nbr\_ids}). 
The data position is represented by a tuple, with the second item indicating the slot offset in the corresponding vector data block.

\sstitle{Vector Data Blocks}
Figure~\ref{fig:vectorfs}(d) depicts the vector data block,
raw vectors are placed in fixed-sized slots.

\subsubsection{Runtime Behaviors}\label{sec:fs-runtime}
We illustrate the runtime behavior of the vector file system.

\stitle{Booting} 
The vector file system boots by loading the meta block to retrieve metadata, followed by loading the bitmap blocks and schema blocks into memory.

\stitle{Opening} 
A file is identified by its name through the schema blocks, which reference the initial PQ and vector mapping blocks. All subsequent blocks are loaded into memory sequentially via their linked block numbers. As in existing on-disk graph-based vector systems~\cite{jayaram2019diskann, wang2024starling, yu2025topology, xu2023spfresh}, data in these blocks is small and often loaded into memory prior to access.

\stitle{Locating}
Given a vector ID, it first determines the vector index position as a tuple (\textsf{PageID}, \textsf{offset}) by the contents in the vector mapping block, which is loaded in memory when opening. 
This tuple is then used to locate the corresponding slot in the vector index block and find the vector position by its \textsf{dpos}, as the {\color{red}} arrows shown in Figure~\ref{fig:vectorfs}.

\stitle{Modifying} 
Modifying a block is done by our page access protocol in Section~\ref{sec:buffer}.
Specifically, the bitmap in the bitmap blocks manages block allocation and recycling, while the bitmaps in the vector index and data blocks enable slot reuse to minimize fragmentation.

\stitle{Effectiveness Analysis} With our vector file system, vector updates need to access only the vector index blocks. Moreover, because these blocks are more compact, they are more likely to remain cached in our buffer manager. Consequently, this approach effectively addresses Issue 4 in Section~\ref{sec:challenges}.

\section{Experimental Evaluation}\label{sec:exp}

In this section, we present the experimental setting in Section~\ref{sec:setting}, then evaluate the performance of \sysname{} to answer the following questions:
\squishlist
\item How does \sysname{} compare to other state-of-the-art vector index update solutions? (Section~\ref{sec:overall})
\item How does the vector-level update perform under different update scenarios? (Section~\ref{sec:effectiveness})
\item How does each system component contribute to the overall performance in \sysname{}? (Section~\ref{sec:breakdown})
\squishend


\subsection{Experimental Setting} \label{sec:setting}

\begin{figure*}[t]
  \centering
  \includegraphics[width=2.0\columnwidth]{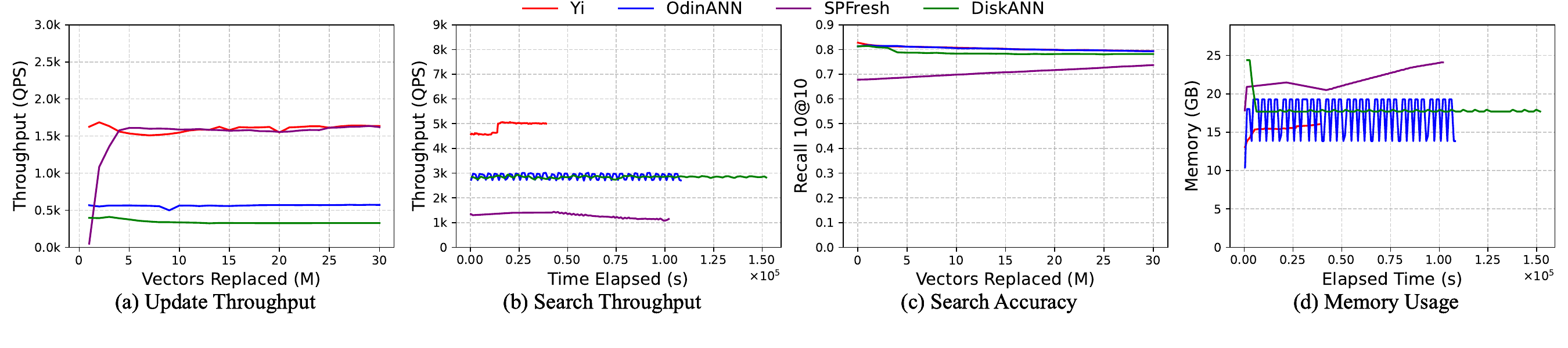}
  \trim \trim 
  \caption{Overall performance in DEEP100M dataset, inserting and deleting 30M vectors from an index of 100M.}
  \label{fig:overall_deep100m}
  \trim \trim 
\end{figure*}

\sstitle{Hardware Platform} All experiments are run on a machine with 2.9GHz INTEL(R) XEON(R) GOLD 6542Y processors. The SSD is 7TB INTEL SSDPF2KX076T1.

\sstitle{Tested Datasets} we use the widely-used vector datasets by following~\cite{xu2023spfresh,guo2025odinann}. Their descriptions are as follows.
\squishlist
\item \sstitle{SIFT~\cite{sift, jegou2011searching}} It is a classic image dataset with 1 billion 128D byte vectors and 10,000 queries.
\item \sstitle{DEEP~\cite{nips2022data}} It consists of image descriptors with 1 billion 96D float vectors and 10,000 queries.
\squishend
We adopt a naming convention where DEEP100M refers to the first 100M vectors in the DEEP dataset~\cite{guo2025odinann}.

\sstitle{Compared Systems}
We compare \sysname{} with the state-of-the-art vector update systems in both cluster-based and graph-based categories. 

\squishlist
\item \stitle{DiskANN~\cite{singh2021freshdiskann}} It is an out-of-place graph-based vector index update system. In particular, DiskANN buffers all insertions and deletions temporarily in memory and merges them into disk when the number of vectors reaches a certain threshold.

\item \stitle{OdinANN~\cite{guo2025odinann}} It is the state-of-the-art out-of-place graph-based vector index update system. In particular, OdinANN adopts direct insertion for new vectors and relies on a full-scan merge process to handle deletions.

\item \stitle{SPFresh~\cite{spfreshgithub}} It is the state-of-the-art in-place cluster-based vector index update system. In particular, SPFresh introduces a partition rebalancing protocol, LIRE, to split or merge clusters for in-place insertions and deletions.
\squishend

\subsection{Overall Performance}\label{sec:overall}

In this section, we compare the performance of \sysname{}, OdinANN, and SPFresh in four metrics: \emph{update throughput, search throughput, search accuracy}, and \emph{memory usage}.
Following the experimental setting of OdinANN, search queries are issued concurrently in a given frequency during vector updates, and search accuracy is evaluated under a fixed search parameter. 
The thread allocation of the compared systems is listed below. The index settings for SPFresh  follow its default configuration.

\squishlist
\item \sstitle{DiskANN} 24 threads in total. 12 threads for search, 12 for its offline merge process.
\item \sstitle{OdinANN} 24 threads in total. 12 threads for search, 12 for its online insert and offline merge.
\item \sstitle{SPFresh} 12 threads in total, where 6 threads of them are used by default configuration~\cite{xu2023spfresh, spfreshgithub}, and the rest 6 threads are dedicated to be search threads to maximize search throughput.
\item \sstitle{\sysname{}} 12 worker threads in total, each thread can process both search and update queries in \sysname{} due to the design of our \tasklet{}-based execution engine (see Figure~\ref{fig:generation}). It is worth pointing out that the number of used threads of \sysname{} is fewer than those of DiskANN and OdinANN.
\squishend

\sstitle{Other Relevant Parameters} \sysname{} uses a 4GB in-memory buffer (Section~\ref{sec:buffer}), and the size of its LRU-based delete list (Section~\ref{sec:vector-level}) is set to 4\% of the total number of vectors in the initial index. For all graph-based indexes (i.e., DiskANN, OdinANN, and \sysname{}), following OdinANN's experimental setup, we set the maximum on-disk out-degree parameter $R$ to 96. For the cluster-based index SPFresh, we set the maximum cluster size to 48KB for the DEEP dataset, consistent with its open-source implementation.

\subsubsection{Performance Evaluation on DEEP100M}\label{sec:deep100m}
In this experiment, we perform 30M insertions and 30M deletions on each system's initial index built on DEEP100M. When updating, the batch size for DiskANN and OdinANN is set to 1M (i.e., 1\% of the data size).

Figure~\ref{fig:overall_deep100m} presents the results for all compared systems. \sysname{} consistently outperforms all baselines across the evaluated metrics. We discuss the results for each metric in detail below.


\stitle{Update Throughput} 
As shown in Figure~\ref{fig:overall_deep100m}(a), \sysname{} maintains a stable update throughput of around 1.6K QPS throughout the 30M replacements. Overall, \sysname{} achieves 2.8$\times$ higher update throughput than OdinANN and 4.7$\times$ higher than DiskANN. SPFresh reaches a similar update rate after warming up, but it comes with much lower search throughput and accuracy, which we will describe later.

\stitle{Search Throughput} 
Figure~\ref{fig:overall_deep100m}(b) shows that \sysname{} provides the highest concurrent search throughput during updates. \sysname{} maintains about 4.8K QPS on average, compared with 2.9K QPS for OdinANN, 2.8K QPS for DiskANN, and 1.2K QPS for SPFresh. This corresponds to 1.7$\times$, 1.7$\times$, and 3.9$\times$ higher search throughput, respectively. In addition, out-of-place update systems exhibit fluctuating search throughput due to their varying CPU and I/O resource demands between the online and offline phases.

\stitle{Search Accuracy} Figure~\ref{fig:overall_deep100m}(c) reports the recall of all compared systems during the replacement of 30M vectors. First, \sysname{} achieves recall nearly identical to OdinANN and consistently higher than DiskANN, maintaining high recall throughout the experiment. This result validates the effectiveness of the vector-level update mechanism in \sysname{}. Second, SPFresh starts with a lower recall (67.8\%) due to its coarse-grained, cluster-based index design, but its LIRE protocol gradually improves recall as clusters adapt to the updated data.
Still, SPFresh cannot deliver high throughput and high recall at the same time.

\stitle{Memory Usage}
Figure~\ref{fig:overall_deep100m}(d) shows that \sysname{} also has the lowest memory footprint. Its memory usage stays around 15.4GB during the run, lower than OdinANN, SPFresh, and DiskANN. \sysname{} also avoids the large memory spikes and oscillations observed in OdinANN, keeping memory usage stable while sustaining higher update and search throughput.


\begin{figure*}[t]
  \centering
  \includegraphics[width=2.0\columnwidth]{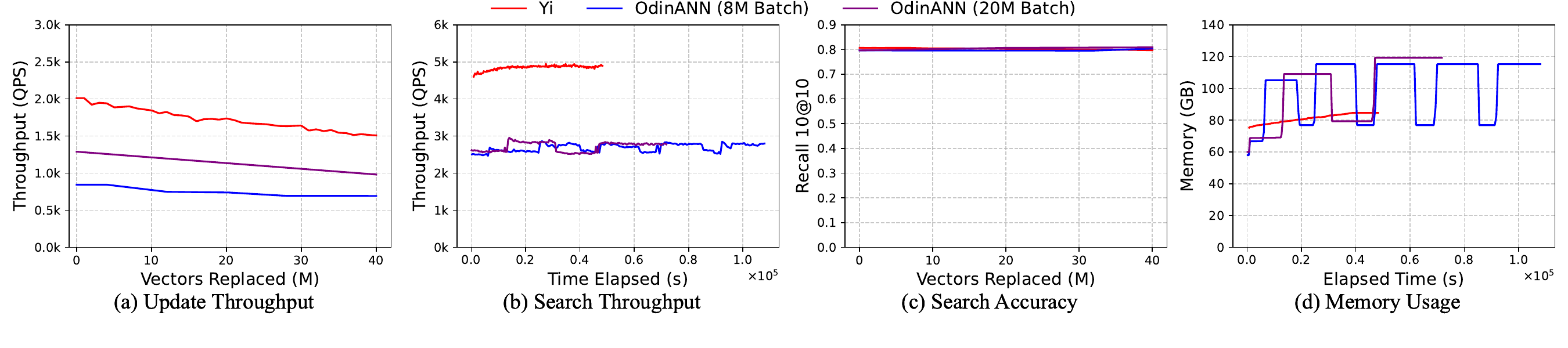}
  \trim \trim
  \caption{Overall performance in SIFT800M dataset, inserting and deleting 40M vectors from an index of 800M.}
  \label{fig:overall_deep800m}
  \trim \trim 
\end{figure*}

\subsubsection{Performance Evaluation on SIFT800M}
In this experiment, we scale the dataset size to 800M. Specifically, we perform 40M insertions and 40M deletions on each system's initial index built on SIFT800M. Due to the low efficiency of DiskANN and SPFresh observed in previous experiments, we omit them from this evaluation and mainly focus on comparing OdinANN under different batch-size settings, as described below.

To evaluate the impact of batch size on out-of-place graph-based vector indexes, we consider two batch sizes: 8M (1\% of the initial size) and 20M (2.5\% of the initial size). As shown in Figure~\ref{fig:overall_deep800m}, \sysname{} consistently outperforms OdinANN under both batch-size settings in terms of update throughput and search throughput while maintaining comparable recall. In particular, \sysname{} achieves 2.3$\times$ and 1.5$\times$ higher average update throughput than OdinANN with 8M and 20M batch sizes, respectively. For search throughput, \sysname{} achieves about 1.8$\times$ higher average throughput than OdinANN under both batch-size settings.
Moreover, as described in Section~\ref{sec:deep100m}, OdinANN exhibits noticeable search-throughput fluctuations across different stages of the update process.

As shown in Figure~\ref{fig:overall_deep800m}(d), \sysname{} consumes less memory than OdinANN under both batch-size settings. Specifically, the peak memory usage of \sysname{} is only 73\% and 71\% of OdinANN's peak memory usage with 8M and 20M batches, respectively. This reduction is mainly because OdinANN requires substantial additional memory during merge operations on SIFT800M, whereas \sysname{} only maintains a fixed 4GB update buffer.

These results demonstrate the efficiency of \sysname{} for large-scale vector updates by avoiding the costly periodic offline merge operations required in out-of-place update systems.

\subsection{Effectiveness of Vector-level Update}\label{sec:effectiveness}
In this section, we evaluate the effectiveness of the vector-level update mechanism in \sysname{} from multiple perspectives. Specifically, we first compare our update mechanism with the existing in-place update algorithm IP-DiskANN~\cite{xu2025place}. We then evaluate its recall performance under different update scenarios, including 100\% replacement, deletion-only workloads, and updates with spatial locality.


\begin{figure}
  \small
  \centering
  \begin{tabular}{cc}
   \multicolumn{2}{c}{
   \includegraphics[width=0.7\columnwidth]{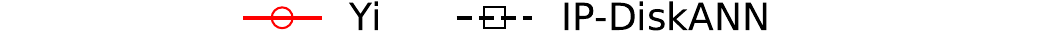}}
   \\

   \subcaptionbox{Update throughput\label{fig:ipdiskann:a}}
   {\includegraphics[width=0.47\columnwidth]
   {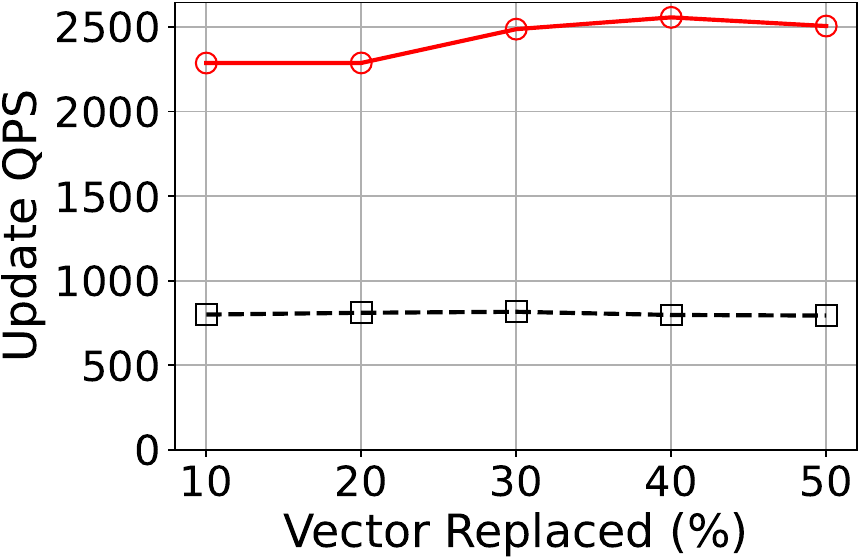}}
   &
   \subcaptionbox{Recall\label{fig:ipdiskann:b}}
   {\includegraphics[width=0.47\columnwidth]
   {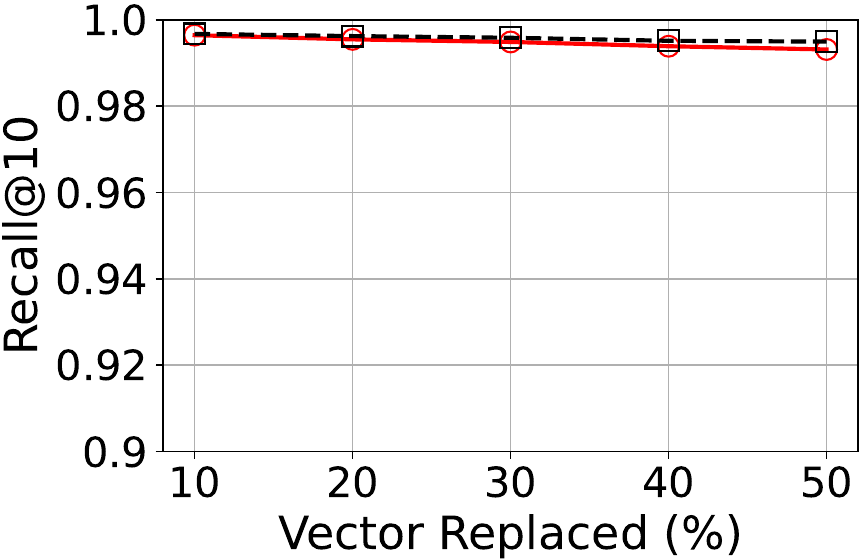}}
  \end{tabular}

  \trim \trim
  \caption{Evaluation of \sysname{} and IP-DiskANN.}
  \label{fig:ipdiskann}
  \trim \trim
\end{figure}


\sstitle{Compare with IP-DiskANN} In this experiment, we first implement the heuristic delete algorithm in IP-DiskANN based on the system components of \sysname{} and compare its performance with our \sysname{} by using its default parameter setting.
We build a graph index of DEEP10M, then replace 5M vectors from it for each method with 12 cores.
Figures~\ref{fig:ipdiskann}(a) and~\ref{fig:ipdiskann}(b) show the update throughput and recall of both methods under the same search parameter during updates. As shown in the figures, our update mechanism achieves recall comparable to that of IP-DiskANN while delivering 3$\times$ higher update throughput. This improvement mainly comes from consolidating the connection-establishment steps across multiple updates, whereas IP-DiskANN incurs excessive random vector accesses during its update process.


\begin{figure}
  \small
  \centering
  \begin{tabular}{cc}
   \multicolumn{2}{c}{
   \includegraphics[width=0.7\columnwidth]{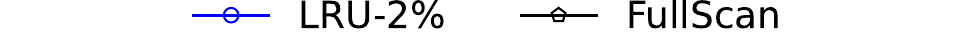}}
   \\

   \subcaptionbox{DEEP1M}
   {\includegraphics[width=0.47\columnwidth]
   {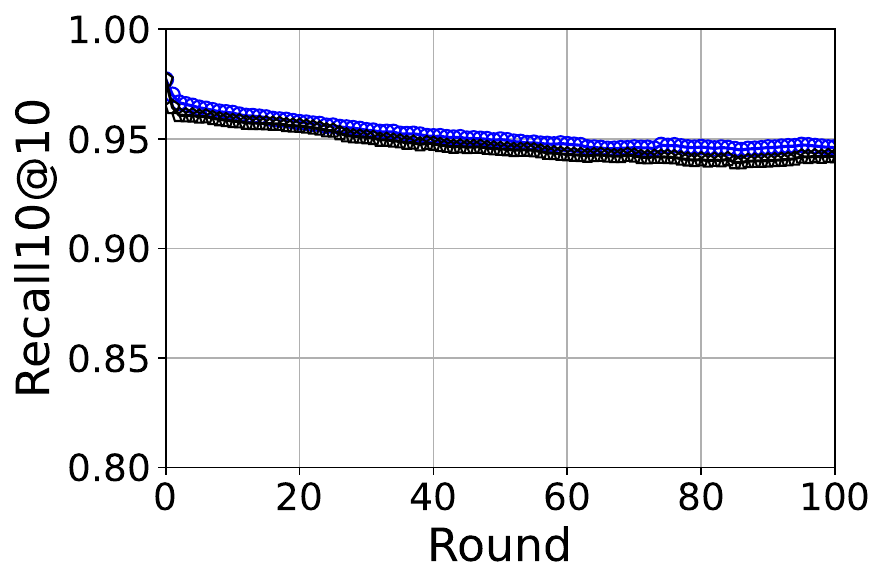}}
   &
   \subcaptionbox{SIFT1M}
   {\includegraphics[width=0.47\columnwidth]
   {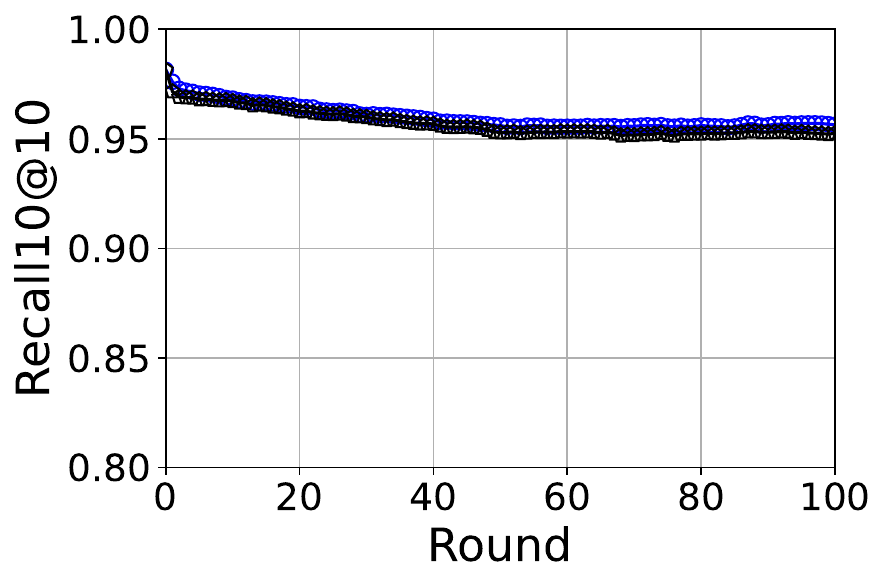}}
  \end{tabular}

  \trim \trim
  \caption{Recall of \sysname{} and full scan merge for 100\% replacement.}
  \label{fig:ins_del}
  \trim \trim
\end{figure}

\begin{figure}
  \small
  \centering
  \begin{tabular}{cc}
   \multicolumn{2}{c}{
   \includegraphics[width=0.7\columnwidth]{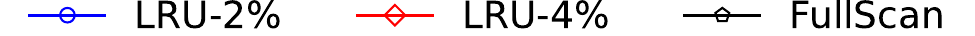}}
   \\

   \subcaptionbox{DEEP1M}
   {\includegraphics[width=0.47\columnwidth]
   {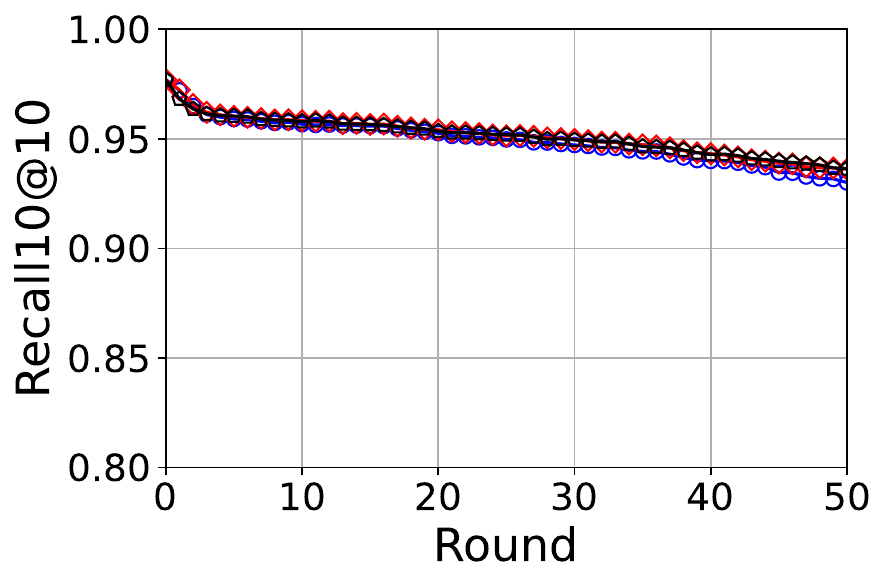}}
   &
   \subcaptionbox{SIFT1M}
   {\includegraphics[width=0.47\columnwidth]
   {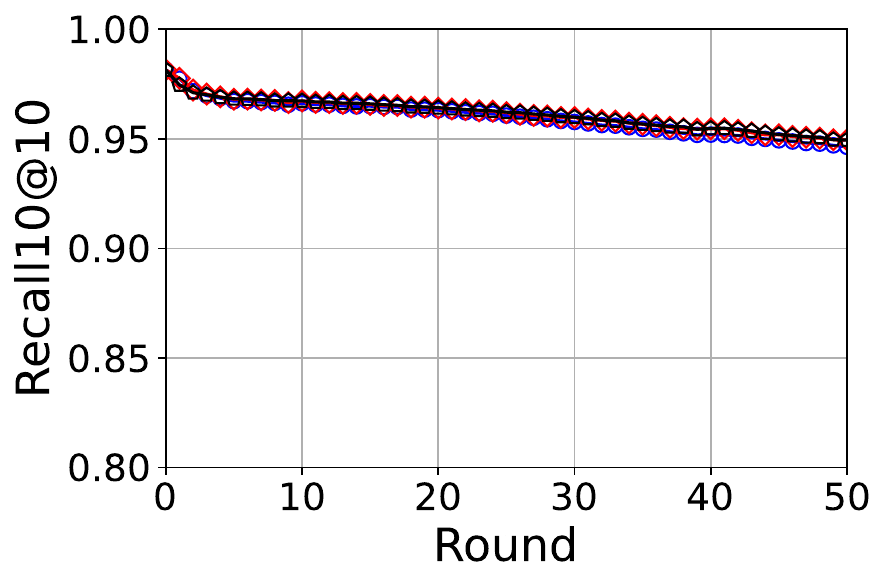}}
  \end{tabular}

  \trim \trim
  \caption{Recall of \sysname{} and full scan merge for 50\% delete only.}
  \label{fig:delete50}
  \trim \trim
\end{figure}


\sstitle{100\% Replacement} In this scenario, we repeatedly delete and insert 10K points in the initial graph index for DEEP1M and SIFT1M datasets in 100 rounds. LRU-2\% employs a 2\%-sized LRU-based delete list (i.e., 20K) and only establishes the connections by the \textsf{connect} tasks generated by the inserted vectors. FullScan performs a full scan for deletions in each round.

As shown in Figure~\ref{fig:ins_del}, LRU-2\% achieves slightly higher recall than FullScan. Moreover, the recall of both methods gradually converges by the end of the experiment, further confirming that our update mechanism can effectively preserve graph quality while avoiding the overhead of full scans.

\sstitle{Deletion Only} We evaluate our vector-level update mechanism under deletion-only scenarios. Specifically, we compare our update mechanism with a full-scan-based method by examining changes in recall.

We conduct deletion-only experiments on both SIFT and DEEP. For each dataset, 
we build a graph index on 1M vectors and measure Recall@10 under the same initial search setting. We then remove 50\% of the vectors, deleting 10K vectors in each round; FullScan
performs a full scan for deletions in each round.

To trigger the \textsf{connect} tasks, we randomly select a set of vectors whose size equals the graph out-degree, matching the number of tasks that would normally be triggered by insertions. Interestingly, as shown in Figure~\ref{fig:delete50}, our recall is slightly lower than the full-scan baseline under the same 2\% LRU setting used in the 100\% replacement experiment. However, this gap can be easily narrowed by increasing the LRU size.
For example, increasing the LRU to 4\% provides a comparable recall to the full scan, as illustrated by LRU-4\% in Figure~\ref{fig:delete50}. Another interesting observation is that both of the methods will have a recall drop, which indicates that insertions play a crucial role in preserving graph connectivity.

In summary, our update mechanism also performs effectively under deletion-only workloads, achieving recall comparable to that of the full-scan-based approach.

\sstitle{Updates with Spatial Locality} We evaluate the effectiveness of our vector-level update mechanism under update scenarios with spatial locality by comparing its recall with that of a full-scan-based method.

\begin{figure}
  \small
  \centering
  \begin{tabular}{cc}
   \multicolumn{2}{c}{
   \includegraphics[width=0.7\columnwidth]{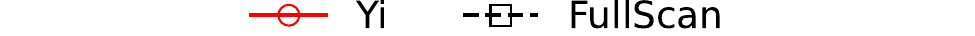}}
   \\[-4pt]
   \subcaptionbox{DEEP1M}
   {\includegraphics[width=0.47\columnwidth]
   {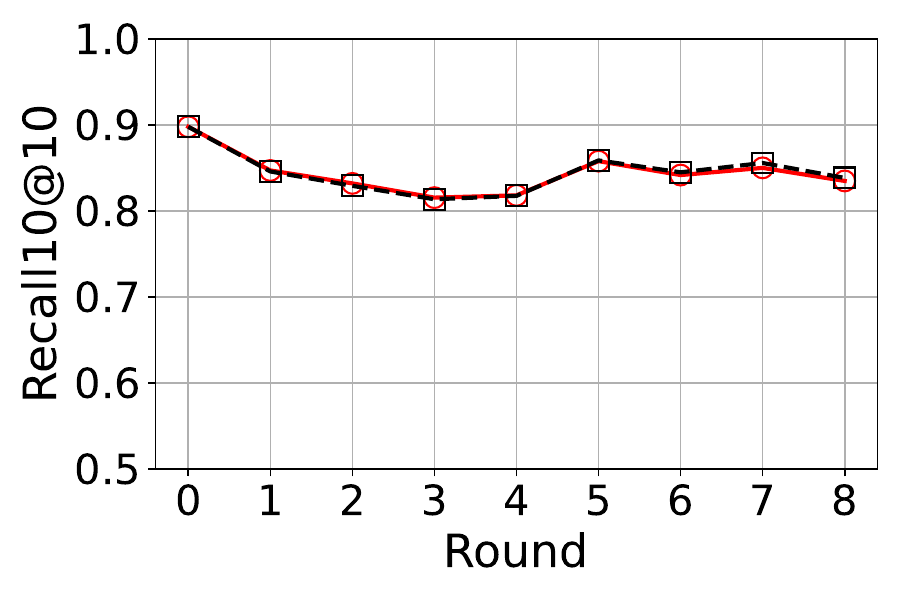}}
   &
   \subcaptionbox{SIFT1M}
   {\includegraphics[width=0.47\columnwidth]
   {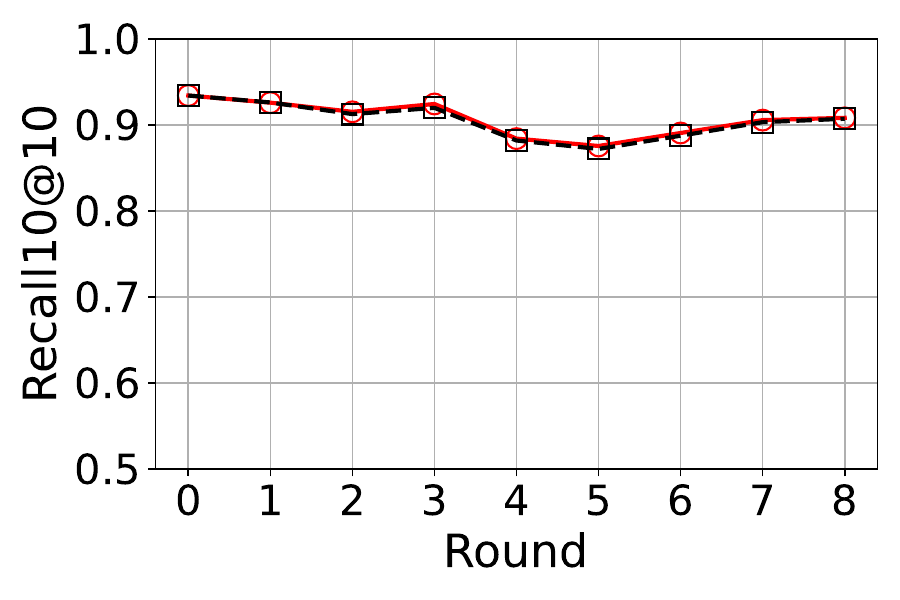}}
  \end{tabular}

  \trim \trim
  \caption{Recall comparison between \sysname{} and a full-scan-based merge method under update scenarios with spatial locality.}
  \label{fig:cluster_recall}
  \trim \trim
\end{figure}

To introduce spatial locality, we first extract 1M vectors from a dataset and apply K-Means to partition them into 16 clusters, denoted as $c_1,\ldots,c_{16}$. These clusters vary in both size and spatial distribution. We initialize the index using the vectors from first eight clusters, i.e., $c_1,c_2,\ldots,c_8$. Then, in round $i$ ($i \leq 8$), we delete vectors from cluster $c_i$ and insert vectors from cluster $c_{i+8}$, thereby maintaining a constant number of eight active clusters throughout the experiment. After each round, we measure the change in recall. We conduct this experiment on both the SIFT and DEEP datasets. For the full-scan-based method, a merge operation is triggered at the end of each round.

As shown in Figure~\ref{fig:cluster_recall}, our update mechanism achieves accuracy comparable to that of the full-scan-based method. Interestingly, unlike the recall trends observed in previous experiments, the recall of both methods is less stable in this setting. This behavior is likely caused by the insertion of out-of-distribution vectors, which can degrade graph quality because search results for such vectors tend to be less accurate. This observation suggests a potential direction for future improvements.

Overall, the results show that \sysname{} can effectively preserve graph quality even when updates exhibit clustered spatial locality.

\subsection{Breakdown Analysis} \label{sec:breakdown}

In this section, we conduct experiments to study the contribution of each key component in \sysname{}. Specifically, we first disable all the components, then we gradually enable each of them to quantify the contribution of them.

We first build an index with the DEEP10M dataset and then perform 100K vector insertions and deletions.
In particular, we use 12 cores and 4GB buffers in our buffer manager.

\begin{figure}[t]
  \small
  \centering
  \includegraphics[width=0.70\columnwidth]{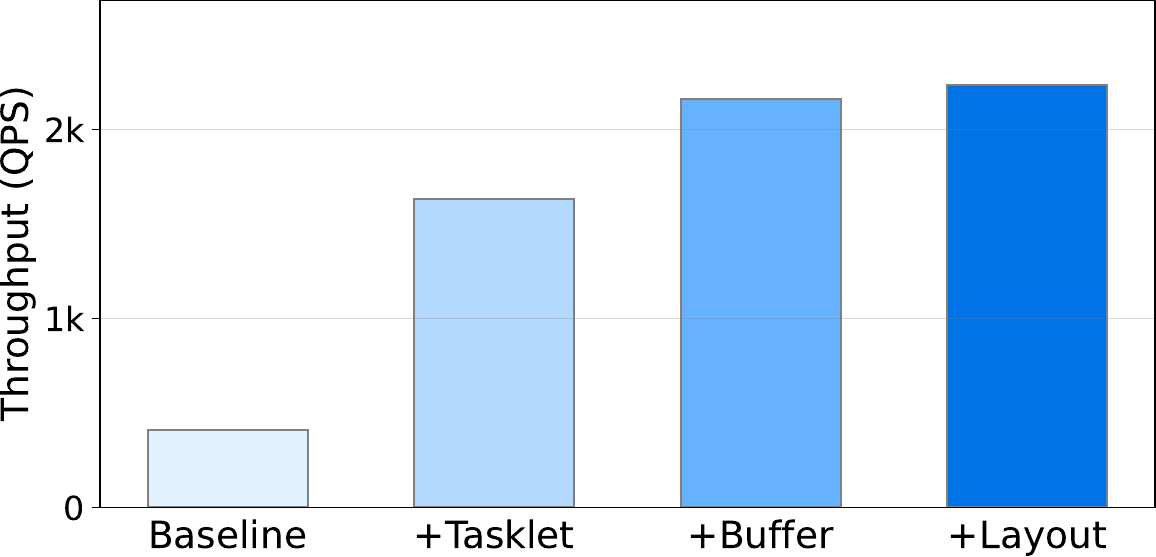}
  \trim \trim 
  \caption{Breakdown Analysis.}
  \label{fig:breakdown}
  \trim \trim
\end{figure}

\stitle{Baseline} We construct a Baseline by disabling the core innovations of \sysname{}. Specifically, we (1) replace the suspending mechanism in our \tasklet{}-based execution engine in Figure~\ref{fig:generation} with synchronous waiting; (2) disable cache persistence in our buffer manager, causing immediate eviction of unreferenced buffers; and (3) adopt a coupled layout.

\stitle{+\Tasklet{}} We generate \tasklet{} by suspending I/O operations and spawning subtasks as in Figure~\ref{fig:generation}(a) and Figure~\ref{fig:generation}(c).
It boosts update throughput by 4.02x by resolving Issue 2 in Section~\ref{sec:challenges}.
In particular, if \tasklet{}-based execution is turned off, the concurrency control mechanism in Section~\ref{sec:buffer} degenerates to the traditional lock-coupling approach.

\stitle{+Buffer} We persist the buffers in the free list or the dirty list in our buffer manager, as in Figure~\ref{fig:breakdown}.
It boosts update throughput by 1.33x by resolving Issue 3 in Section~\ref{sec:challenges}.

\stitle{+Layout} We follow the vector file system layout shown in Figure~\ref{fig:vectorfs}.
This design improves update throughput by 1.04× and increases the cache hit ratio from 66\% to 78\%.










\section{Related Work}\label{sec:related}

\stitle{Vector Databases}
Systems like PASE~\cite{yang2020pase}, AnalyticDB-V~\cite{wei2020analyticdb}, SingleStore-V~\cite{chen2024singlestorev}, Milvus~\cite{milvus2021}, pgvector~\cite{pgvector25}, and Vexless~\cite{su2024vexless} support vector search via database or interfaces. Most adopt DiskANN-style graph indexes but rely on rebuild-based updates~\cite{singh2021freshdiskann}, incurring high overhead. \sysname{} introduces an in-place update paradigm that enables efficient, concurrent vector updates while maintaining search performance.

\stitle{Dynamic Graph Processing} Systems such as LLAMA~\cite{macko2015llama}, GraphOne~\cite{khayyat2017graphone}, and Aspen~\cite{spear2020aspen} support dynamic graph updates via multi-versioning or append-only designs. However, they are not suitable for graph-based vector indexes: ANN graphs require bounded degrees for search efficiency, and their concurrency control targets graph analytics, not ANN search. \sysname{} proposes a dedicated architecture for in-place updates on ANN graphs.

\section{Conclusion}\label{sec:con}
In this paper, we present \sysname{}, an efficient in-place graph-based vector update system. We first analyze the limitations of out-of-place solutions and the challenges of supporting in-place updates. 
Guided by the principle that decomposition facilitates consolidation, 
we propose a vector-level update mechanism and implement it on \sysname{} with three core components: (i) a \tasklet{}-based execution engine, (ii) an asynchronous buffer manager, and (iii) a vector file system. Extensive experiments on real-world datasets show that \sysname{} substantially outperforms state-of-the-art methods.
In future work, we plan to (a) leverage SPDK to accelerate the vector file system of \sysname{}, and (b) accelerate the performance of \sysname{} in multi-tenant scenarios.

\bibliographystyle{ACM-Reference-Format}
\bibliography{ref}

\end{document}